\documentclass[aps,pra,preprint,floatfix]{revtex4}
\usepackage{amssymb, amsmath}
\usepackage{color}      
\usepackage{graphicx}	
\usepackage{subfigure}
\usepackage{float}
\usepackage{multirow}
\usepackage{stackrel}
\usepackage{mathtools}
\DeclareGraphicsExtensions{.eps,.png,.pdf,.jpg}

\def\eeq{\relax}
\def\beq#1#2\eeq{\begin{equation}\label{#1}#2\end{equation}}
\def\bal#1#2\eal{\begin{align}\label{#1}#2\end{align}}
\def\bse#1#2\ese{\begin{subequations}\label{#1}#2\end{subequations}}
\def\ba{\begin{aligned}}   \def\ea{\end{aligned}}

\newcommand{\spring}{- \!{\scriptscriptstyle \backslash \! / \! \backslash \! / \! \backslash \! /} \! -\!}
\newcommand{\hlin}{ \!\!\!\!\! -\!\!\!}

\def\dd{\operatorname{d}} 
\def\ii{\operatorname{i}}

\def\Re{\operatorname{Re}} 
\def\sgn{\operatorname{sgn}} 

\begin{document}

\title{A broadband solid impedance transformer for acoustic transmission between water and air}

\author{Hesam Bakhtiary Yekta and Andrew N. Norris} 

\affiliation{Department of Mechanical \& Aerospace Engineering, Rutgers University,
98 Brett Road, {Piscataway},       NJ    {08854},            {USA}}

\begin{abstract}

{Total acoustic transmission between air and water  was shown in our recent paper to be attainable with  a solid interface  comprising two  parallel thin elastic plates connected by rigid ribs, although the transmissivity  is a  narrow-band effect.}   We demonstrate  here that  broadband transmission  can be obtained by introducing a third, central plate.  A theoretical analysis combined with numerical optimization shows that the optimal 3-plate impedance transformer has a central plate far thicker than the others. This implies a simpler interpretation of the optimal 3-plate impedance transformer as two elastic plates separated by  a mass-like impedance.  The characteristics of the broadband transformer {may then}  be understood using results for the {previously studied} 2-plate system and asymptotic approximations using the small air-to-water impedance ratio.  Optimal systems with  water and air-side plates of similar material have relative thicknesses of approximately three to one, respectively, with the central mass having areal density approximately  17 times the water side plate.  Further identities relate  the frequency of total transmission to the plate thicknesses and to the rib separation length.    The impedance transformer is compared to an ideal two layer quarter wavelength model, allowing us to identify a minimal attainable Q-factor of about 5.5, which is achieved in examples presented. 
{The  formulas for approximately optimized parameters also serve as the initial population for numerical optimization, greatly  accelerating the process.} 
Together, the theoretical and numerical results point to a remarkably simple class of purely solid impedance transformers, with system parameters well defined by the asymptotically small parameter: the ratio of air-to-water acoustic impedances. 
\end{abstract}

\maketitle



\section{Introduction} \label{sec1}   

Transmission of acoustic, or other wave motion, between dissimilar media can be achieved with appropriate impedance matching.  The challenge in sending sound  from  air to  water, or vice versa, is to   transmit  between two highly dissimilar materials  with an impedance ratio in excess of 3,600.   Despite the difficulty, several approaches have been proposed, beginning with 
Bok et al.\ \cite{Bok2018} using an air layer as a spring combined with a membrane mass in series. Other   solutions are of this type, with a membrane or fluid layer acting as a mass  and an air layer acting as a stiffness, together making  a sub-wavelength resonator.  Huang et al.\  \cite{Huang2021a} proposed a hydrophobic structure  rather than a membrane to separate water and air. 
Other methods include bare bubbles  \cite{Cai2019,Lee2020}, useful  for their acoustic sub-wavelength resonating features. Gong et al.\ \cite{Gong2023} used polyester membranes to separate the air bubbles from water and studied the effect of membrane viscosity on the transmitted energy. Liu et al. \cite{Liu2023} employed an air-channel mechanism that is placed in the interface of air and water to provide the impedance matching. Zhou et al.\ \cite{Zhou2023,Zhang2024} designed a layer of 3D printed epoxy  to create an impedance matching between water and air, while Dong et al.\ \cite{dong2020bioinspired} used a bioinspired metagel impedance transformer to overcome narrow bandwidth limits. A gradient index matching layer that combines air-based and water-based metafluids was demonstrated by Zhou et al.\ \cite{Zhou2023a}.
Near-perfect air-water transmission can also be achieved, in principle, with  a gradient Willis-like acoustic metamaterial 
\cite{Tian2024}.

An alternative approach to total acoustic transmission between water and air was recently proposed \cite{HBYANN2025} {using  an interface of} two elastic plates separated by periodically spaced  ribs.   The model is  all solid, e.g.\ aluminum, requiring no interfaces between water and air, and in particular, it allows for asymptotic analysis based on the small parameter defined by the impedance ratio {of air to water}.  This leads to several results, such as   that the  lower bound for the Q-factor is  $30.59$, which  is simply related to the water-air impedance ratio.  The "flex-layer" transformer of \cite{HBYANN2025} is  a sub-wavelength metamaterial realization of  the classical Hansell \cite{Hansell}  quarter-wavelength intermediate layer with an impedance equal to the harmonic mean of the two media.  

The purpose of this paper is to provide a broadband version of the flex-layer  \cite{HBYANN2025}.   The proposed model adds a central plate to the 2-plate flex-layer, although  it is found {from numerical optimization of the system parameters that optimal transmission is obtained if the central plate is very thick, in which case it acts as a mass}.  This observation leads to considerable simplification and   asymptotic approximations that provide accurate initial optimal designs in terms of bandwidth and transmittivity.   It also allows us to show that the present model is a sub-wavelength  realization of Hansell's  two-layer quarter-wavelength solution,   where each layer's impedance is the harmonic mean of its neighboring  {impedances}.   This explains and quantifies the large frequency bandwidth, an order of magnitude greater than our previously proposed  2-plate design  \cite{HBYANN2025}. 

The impedance transformer model is introduced in Section \ref{sec2} along with the general solution for plane wave transmission and reflection. Numerical examples in  Section \ref{sec3} indicate that broadband near-total transmission is possible with specific combinations of the four length parameters that define the transformer model.   The remainder of the paper explains the physical basis for these optimal model parameters. The starting point is an observation from the numerical examples that optimal transformers have a central plate far thicker than the other two, leading to an approximate but accurate model explored in Section \ref{sec4}. Using asymptotic analysis based on the small parameter defined by the air-water impedance ratio, several important identities are obtained linking model dimensions.  These are explained in  Section \ref{sec5} in terms of a pair of coupled resonators defined by elements of the 3-plate system.  A further identity is found by comparing the resonators with a simpler spring-mass model that has explicit solution.  A summary of the main results is given in   Section \ref{sec6}.

\section{Full dynamic model of scattering from a flex-layer}  \label{sec2}   

\begin{figure}[h]
    \centering
    \includegraphics[width=0.6\textwidth]{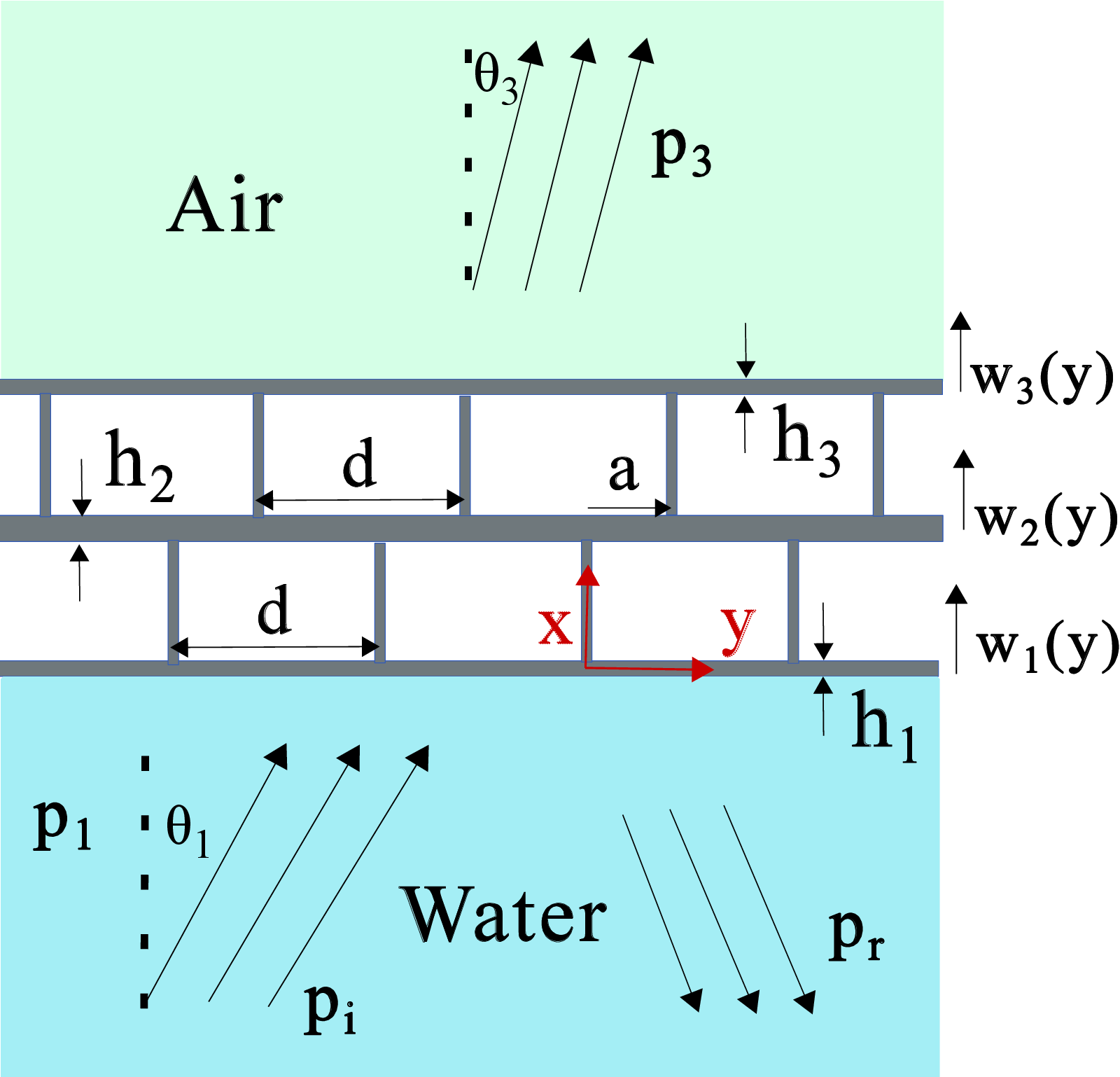}
    \caption{The plates of  the asymmetric panel are separated by ribs set a distance $d$ apart with the intermediate space assumed to be vacuum.  The ribs between plates 2 and 3 are staggered relative to the  ribs between plates 1 and 2 by spacing $a \le \frac d 2$.  The formulation considers a plane wave incident from the water side. }
    \label{Flex_3Plates}
\end{figure}

 We consider time harmonic acoustics with unstated time dependence $e^{-\ii \omega t}$.   A plane wave 
 {$p_i = p_0 \, e^{\ii k_1 (x \cos\theta_1 +y \sin \theta_1)} $}
 is incident from water ({with label 1}) at angle $\theta_1$ from the normal, with $y-$wavenumber $ k_1\sin\theta_1 \equiv k_0$ where $k_j=\omega /c_j$, $j=1,3$, see Fig.\ \ref{Flex_3Plates}.  
The incident acoustic pressure in water along with its rigidly reflected pressure, 
{$p_r = p_0 \, e^{\ii k_1 (-x \cos\theta_1 +y \sin \theta_1)} $, } together  give zero normal velocity  on  plate 1.  The  normal velocity, $v_1(y) = v_x(-0,y)$ is  therefore related to {an  additional pressure $p_1$ which radiates away from the plate, such that the total pressure in water is $p=p_i +p_r +p_1$.}
On the air side ({with label 3}) the total acoustic pressure $p = p_3$  radiates in the positive $x-$direction, with plate normal velocity, $v_3(y) = v_x(+0,y)$.
The total acoustic pressure in water $(x<0)$ and air $(x>0)$
is therefore \cite{HBYANN2025}
\beq{-44}
p(x,y) = \begin{cases} p_1(x,y) +
2p_0  \cos  (k_1 x \cos\theta_1 )\, e^{\ii k_1 y \sin \theta_1 }
    , & x<0, 
    \\
    \\
    p_3(x,y), & x>0 .
\end{cases}
\eeq
{The finite gap between the plates is compressed into the single point $x=0$ for simplicity.}

The pressures $p_1$ and $p_3$ are defined by first 
introducing  $y-$transforms for the normal velocities for {the three plates:}
\beq{7=2}
\hat V_j(\xi) = \int_{-\infty}^\infty v_j(y)  e^{-\ii \xi y} \dd y, 
\quad
v_j(y) = \frac 1{2\pi}  \int_{-\infty}^\infty \hat V_j(\xi)  e^{\ii \xi y} \dd \xi \quad j=1,2,3.
\eeq   
The additional scattered pressure in the water $(j=1)$ and the total pressure in the air  $(j=3)$ are 
\beq{7=3}
p_j(x,y) =  \frac {\sgn x}{2\pi}  \int_{-\infty}^\infty \hat Z_{fj}(\xi) \hat V_j(\xi)  
e^{\ii ( \sqrt{k_j^2-\xi^2}\, |x| + \xi y )}\, \dd \xi, \quad j=1, 3
\eeq
where the fluid  impedances are 
\beq{7=4}
\hat Z_{fj}(\xi)= \frac{\rho_j \omega}{ \sqrt{k_j^2-\xi^2}  } , \quad j=1, 3 .
\eeq
Square roots in Eqs.\  \eqref{7=3} and \eqref{7=4}  are either positive real or positive imaginary.

\subsection{Plate equations}\label{subsec3} 
The displacement in the $x-$direction  of the three  plates  satisfy
\bal{1}
   \mathcal{L}_1 w_1(y) = \ & 
 2p_0  e^{\ii k_0 y } + p_1(0,y)
 \notag \\  & - \Big[   Z_{0+} (v_2+v_1)(y) - Z_{0-} (v_2-v_1) (y)  \Big]
  \sum_{l=-\infty}^\infty \delta(y- ld) ,
\notag 
\\
   \mathcal{L}_2 w_2(y) =& - 
 \Big[   Z_{0+} (v_2+v_1)(y) + Z_{0-} (v_2-v_1) (y)  \Big]
 \sum_{l=-\infty}^\infty \delta(y- ld)  
 \notag \\  &- \Big[   Z_{0+} (v_3+v_2)(y) - Z_{0-} (v_3-v_2) (y)  \Big]
 \sum_{l=-\infty}^\infty \delta(y-a - ld),
\\
   \mathcal{L}_3 w_3(y) =& -p_3(0,y)
    \notag \\  & -  \Big[   Z_{0+} (v_3+v_2)(y) + Z_{0-} (v_3-v_2) (y)  \Big]
 \sum_{l=-\infty}^\infty \delta(y-a -ld)
\notag
\eal
with displacements   $w_j(y) = (-\ii \omega)^{-1}v_j(y)$ and plate equations $ \mathcal{L}_j w(y) =  D_j w''''(y) - m_j \omega^2 w(y) $, $j=1,2,3$.    The mass  per unit area in each is  $m_j = \rho_{sj} h_j $ and the bending stiffness is 
  $D_j = E_jI_j/(1-\nu_j^2)$ where $I_j = \frac{h_j^3}{12}$, $j=1,2$. {$Z_{0+}$ and $Z_{0-}$ are rib impedances, assumed for simplicity to be the same for the {two sets of ribs}.  The precise form of the impedances follows from the rib model considered, e.g. spring with mass, wave bearing structure, etc., see \cite{HBYANN2025} for explicit details and examples.  
  $Z_{0+}$ is a mass-like impedance and it is ignorable ($Z_{0+} \approx$ 0) because we assume that the ribs are light.  
  $Z_{0-}$ is a stiffness-like impedance, and is assumed to be very large, modeling a nearly rigid rib.  This allows us to simplify equations later using the  approximations $ Z_{0+} \to 0$ and $ {1}/{Z_{0-}} \to\, 0$. 
  In the following we first set  $ Z_{0+} \to 0$ in \eqref{1} but retain 
  $Z_{0-}$, taking the rigid  limit later.  }
  

Taking the $\xi$ transform of \eqref{1}, and using the  Poisson summation identity \cite{lin1977sound}
\Big( $\sum_{l=-\infty}^\infty    \delta(y-l d)
 = \frac 1d   \sum_{m=-\infty}^\infty e^{-\ii 2\pi m \frac yd}
$ \Big), gives 
\beq{3}
\begin{aligned}
   \hat V_1(\xi) &=    q_{d}(\xi) \,\hat Y_1(\xi)       + 4\pi p_0 \, \hat Y_1(k_0) \, \delta(\xi-k_0)  , 
   \\ 
   \hat V_2(\xi) &=  -  q_{d}(\xi)  \,\hat Y_{p2}(\xi) + q_{a}(\xi) \,\hat Y_{p2}(\xi)    ,
   \\
    \hat V_3(\xi) &=  - q_{a}(\xi) \,\hat Y_3(\xi)   .
    \end{aligned}
\eeq
with the notation $\hat Y_{p2}(\xi) = 1/\hat Z_{p2}(\xi) $,   $\hat Y_j(\xi)  = \big( \hat Z_{p_j}(\xi) + \hat Z_{f_j}(\xi)  \big)^{-1}$,  $j=1,3$, 
plate impedances
\beq{Zplate}
\hat Z_{p_j}(\xi) = \frac {D_j \xi^4 - m_j \omega^2}{-\ii \omega} , \quad j=1,2,3,
\eeq
and
\beq{q}
   \begin{aligned}
    q_{d}(\xi) &= \frac{Z_{0-}}{d} \sum_{m=-\infty}^\infty    \big( \hat V_2 - \hat V_1  \big)( \xi + \frac{2\pi m}{d}) ,
   \\
   q_{a}(\xi) &= \frac{Z_{0-}}{d} \sum_{m=-\infty}^\infty     \big( \hat V_3 - \hat V_2  \big)( \xi + \frac{2\pi m}{d})\,  e^{\ii m \phi}
   \end{aligned}
\eeq 
with phase angle  $\phi = 2\pi \frac{ a}{d}$.

\subsection{Solution} \label{subsec4} 

It follows from their definitions  that $q_{d}( \xi + \frac{2\pi m}{d}) =  q_{d}(\xi)$ and 
$q_{a}( \xi + \frac{2\pi m}{d} ) =  q_{a} (\xi) \,  e^{-\ii m \phi}$. 
At the same time,  Eqs.\ \eqref{3}  allow us to express $\hat V_j(\xi+ \frac{2\pi m}{d})$, $j=1,2,3$, in terms of 
$ q_{d}(\xi)$ and $ q_a(\xi)$.   Upon substitution back into \eqref{q} we obtain a system of equations for the latter quantities: 
\bal{0=}
\begin{pmatrix}
\frac{d}{Z_{0-}} + S_1  + S_{p2}^{(0)} & -  S_{p2}^{(-\phi)}
\\
 -  S_{p2}^{(\phi)} &  \frac{d}{Z_{0-}} +  S_{p2}^{(0)} +  S_3 
\end{pmatrix}
\begin{pmatrix}
 q_{d}(\xi) \\
  q_{a}(\xi)
\end{pmatrix}  &
\\
= -4\pi p_0 \hat Y_1(k_0)  \sum_{m=-\infty}^\infty  \delta(\xi- \xi_m) &
\begin{pmatrix}
1 \\0
\end{pmatrix}
\eal
where  
\beq{X}
\begin{aligned}
 S_j(\xi)  =&  
 \sum_{m=-\infty}^\infty     \hat Y_j \big(\xi +  \frac{2\pi m}{d}\big),
  \ \ j=1, 3; \\
 \quad 
S_{p2}^{(\alpha)} (\xi)  =&  
 \sum_{m=-\infty}^\infty     \hat Y_{p2} \big(\xi +  \frac{2\pi m}{d}\big)
 \, e^{\ii m \alpha}
 \end{aligned}
\eeq
and
\beq{7=6}
\xi_m = k_0 + \frac{2\pi m}{d} .
\eeq
We make the further  assumption of rigid ribs, $1/Z_{0-} = 0$, so that 
  $q_{d}(\xi)$ and $q_{a}(\xi)$ are
\beq{55}
   \begin{aligned}
 q_{d}(\xi) &=-  4\pi p_0  {\hat Y_1(k_0)} B^{-1}(k_0)\,   \big(  S_{p2}^{(0)}(k_0) +  S_3(k_0) \big)   
 \,  \sum_{m=-\infty}^\infty 
    \delta \big(\xi-\xi_m \big),
    \\
     q_{a}(\xi) &= -    4\pi p_0 {\hat Y_1(k_0)} B^{-1}(k_0)\,    S_{p2}^{(\phi)}(k_0)  
 \,  \sum_{m=-\infty}^\infty 
    \delta \big(\xi-\xi_m \big)\,e^{-\ii m \phi} ,
      \end{aligned}
\eeq
where 
\beq{53}
B(\xi ) = 
  \Big(  ( S_1 + S_{p2}^{(0)} ) ( S_{p2}^{(0)} + S_3 ) - S_{p2}^{(-\phi)}  S_{p2}^{(\phi)} \Big) (\xi)  
\eeq
and we have used the periodic properties
$S_j(\xi +\frac{2\pi n}d)= S_j(\xi)$, $j=1,3$
and $ S_{p2}^{(\alpha)}  (\xi+\frac{2\pi n}d )
= S_{p2}^{(\alpha)}  (\xi) e^{-\ii n \alpha}$, for integer $n$.

\subsection{Reflected and transmitted waves}

Total pressure in the incident water $(x<0)$ and the transmitted medium air $(x>0)$ follows from Eq.\  \eqref{55}  as 
\beq{Ref1Coef}
    p(x,y) =
    \begin{cases} p_0 e^{ \ii k_1(x \cos\theta_1 +y \sin \theta_1)}  \\  +p_0 R(\theta_1)\, e^{\ii k_1(-x \cos\theta_1 +y \sin \theta_1)}
    +p_{1ev}(x,y) , & x<0, 
    \\
    \\
    p_0 \, T(\theta_3)\, e^{\ii k_3(x \cos\theta_3 +y \sin \theta_3)}
    +p_{3ev}(x,y),  & x>0, 
   \end{cases}
\eeq
where
\beq{RefCoef}
\begin{aligned}
   R(\theta_1) =& R_1(\theta_1) +  \big( 1-R_1(\theta_1) \big)\, 
    {\hat Y_1(k_0)} B^{-1}(k_0)\, 
  \big(  S_{p2}^{(0)}(k_0) +  S_3(k_0) \big)    ,
    \\
     T(\theta_3) =&   \big( 1-R_3(\theta_3) \big)\, 
     {\hat Y_1(k_0)} B^{-1}(k_0)\, 
    S_{p2}^{(\phi)}(k_0)     .
\end{aligned}
\eeq
 $R_1$ and  $R_3$ are the reflection coefficients for plane wave incidence on the plates,   
\beq{Ref1Coef+}
    R_j(\theta_j) = \frac{\hat Z_{pj}(k_0) - \hat Z_{fj}(k_0)}{\hat Z_{pj}(k_0) + \hat  Z_{fj}(k_0)} ,   \quad j=1,3, 
\eeq
and the evanescent, or near, fields,  are 
\beq{pev1}
\begin{aligned}
    p_{jev}(x,y) = 
 2p_0 \,  
 \frac{\hat Y_1(k_0)} { B(k_0)}
 \, &\sum_{m \neq 0}
    \hat  Z_{fj}\big( \xi_m \big)  \hat Y_j\big( \xi_m \big)  
     e^{\ii \big( (k_{jx})_m\, |x|  +\xi_m y \big)}
        \ 
   \\  & \times\ \begin{cases}
       \big(  S_{p2}^{(0)}(k_0) +  S_3(k_0) \big)   \,  , & j=1, 
      \\
   S_{p2}^{(\phi)}(k_0)  \,  e^{-\ii m \phi} , & j=3, 
      \end{cases}
\end{aligned}
\eeq
where 
$(k_{jx})_m =  \sqrt{k_j^2 -  \xi_m^2 }$.

\subsection{Conditions for total transmission}\label{subsec44}

In order to find conditions necessary to obtain full transmission, we focus on the reflection coefficient $ R(\theta_1) $, which 
must vanish. Based upon Eq.\ \eqref{RefCoef} it takes the form
\beq{7073}
   R(\theta_1) =  R_1(\theta_1)\, 
   { \big( S_{p2}^{(0)} (k_0) + S_3 (k_0) \big) }{ B^{-1}(k_0)} \, 
   \Gamma_1 (k_0)
\eeq
where 
$ \Gamma_1$ can be expressed 
\beq{7-4}
   \Gamma_1 (k_0 )=   {S_1}' (k_0) +  S_{p2}^{(0)} (k_0)
  - \frac{ S_{p2}^{(-\phi)} (k_0)  S_{p2}^{(\phi)} (k_0) }  {  S_{p2}^{(0)} (k_0) + S_3 (k_0)  } 
    +  \frac 1{\hat Z_{p1}(k_0) - \hat Z_{f1}(k_0)} 
\eeq
with  $ {S_1}' (\xi) =  {S_1} (\xi) -  \hat Y_{1}(\xi) $.  
Note that $ {S_1}'$  and $ S_{p2}^{(0)}$ are pure imaginary while 
$ S_{p2}^{(-\phi)} = - \overline { S_{p2}^{(\phi)} } $.  
Hence, {for normal incidence,}  $k_0=0$, the following quantity must vanish at total transmission:
\beq{5+3}
   \Re  \Gamma_1 (0) =  \frac{\alpha_3^2 Z_3}{Z_3^2 + (\omega m_3)^2} - 
\frac{ Z_1}{Z_1^2 + (\omega m_1)^2}
 \ \ \text{with} \ \ 
\alpha_3  = \left| \frac{ S_{p2}^{(\phi)} (0) } 
{   S_{p2}^{(0)} (0) + S_3 (0)  }
\right| 
\eeq
{where $Z_j = \rho_j c_j$, i.e.\ $Z_1$ and  $Z_3$ are  water and air 
impedances, respectively. 
Equation \eqref{5+3} follows from identities such as 
$\Re S_{p2}^{(0)} =0$,  $\Re  S_3 (0)= \Re 
\big( \hat Z_{p3}(0) + \hat Z_{f3}(0) \big)^{-1}$
 and $\hat Z_{fj}(0) -\hat Z_{pj}(0)   = Z_j+ \ii \omega m_j$ for $j=1,3$.
} 

Considering incidence from the air side, the reflection coefficient is 
\beq{7075}
   R(\theta_3) =  R_3(\theta_3)\,  { \big( S_{p2}^{(0)} (k_0) + S_1 (k_0) \big) }{ B^{-1}(k_0)} \, 
   \Gamma_3 (k_0)
\eeq
where $\Gamma_3$ has a form  analogous to $\Gamma_1$. 
Proceeding as before, we have 
\beq{5+38}
    \Re  \Gamma_3 (0) =  \frac{\alpha_1^2  Z_1}{Z_1^2 + (\omega m_1)^2}
- \frac{ Z_3}{Z_3^2 + (\omega m_3)^2}
\ \ \text{with} \ \ 
\alpha_1  = \left| \frac{ S_{p2}^{(\phi)} (0) } 
{   S_{p2}^{(0)} (0) + S_1 (0)  }
\right| .
\eeq

At total transmission both $ \Gamma_1 (0)$ and $ \Gamma_3 (0)$ vanish, requiring from 
Eqs.\ \eqref{5+3} and \eqref{5+38} that 
 $\alpha_1  \alpha_3 = 1$.   The following pair of conditions are therefore necessary and sufficient for total transmission
 \bse{330}
 \bal{2+7}
\left| \big(  S_1 (0)+ S_{p2}^{(0)} (0)  \big)  \big( S_3 (0) + S_{p2}^{(0)} (0)\big) 
\right| &=  \left|   S_{p2}^{(\phi)} (0) \right|^2, 
 \\
 \left| \frac{ S_1 (0) +S_{p2}^{(0)} (0) } {  S_3 (0) + S_{p2}^{(0)} (0) }\right|
 &= \frac{Z_1}{Z_3} \bigg(
 \frac{Z_3^2 + (\omega m_3)^2} {Z_1^2 + (\omega m_1)^2}
 \bigg) .
\label{555}
\eal\ese
The first is a restatement of   $\alpha_1  \alpha_3 = 1$
while the second expresses the condition 
$\Re  \Gamma_1 (0) =0$ or $\Re  \Gamma_3 (0) =0$.

\section{Numerical examples}\label{sec3}  
As a first step in understanding the impedance transformer model, we perform numerical sweeps over a range of system parameters  to find optimally broadband examples of transmission.     The results indicate that optimal transmission can be achieved with a  simplified system, which is discussed in  subsequent Sections.  

\begin{figure}[H]   
    \centering
    \includegraphics[width=0.4\textwidth]{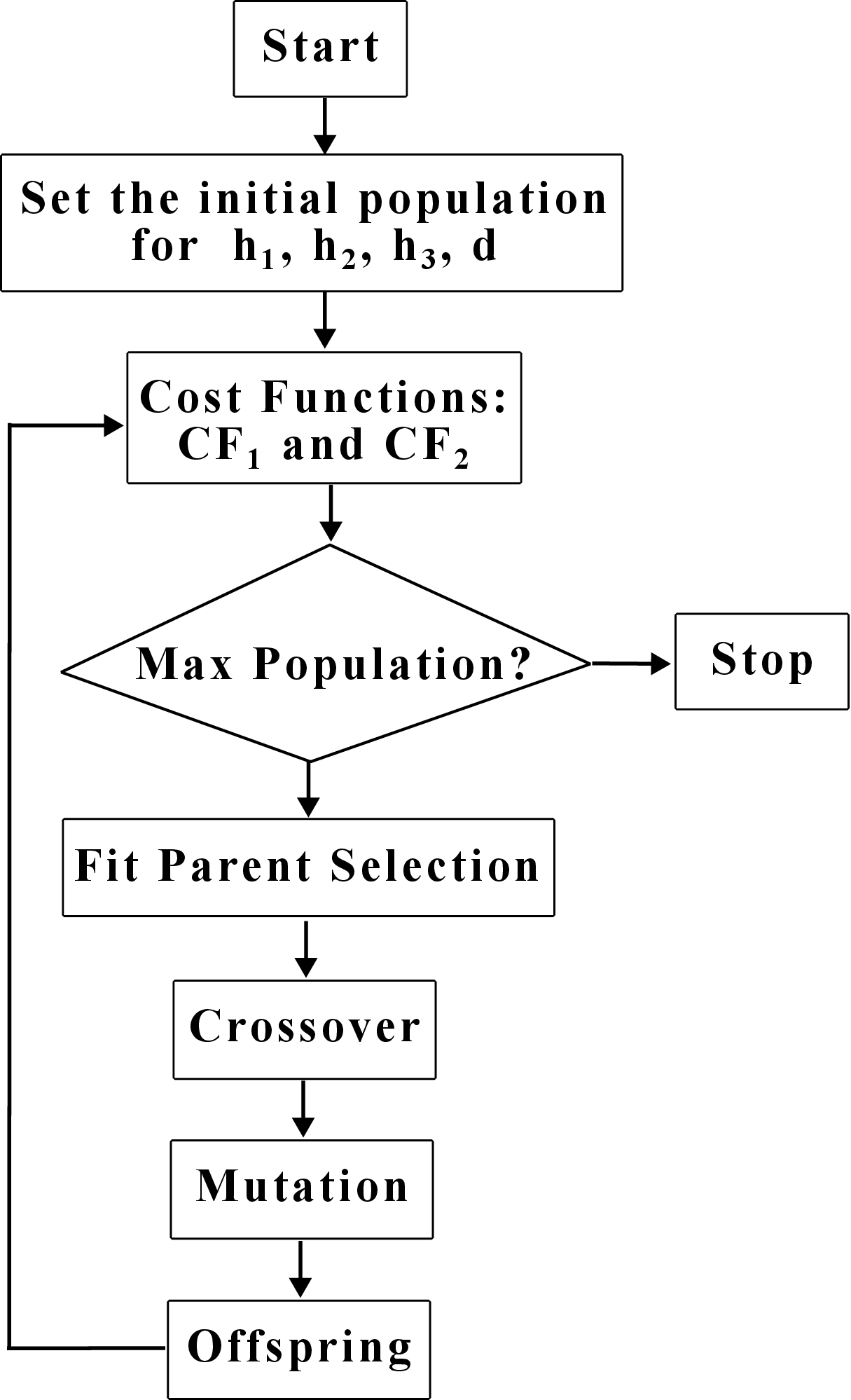}
    \caption{{Optimization flowchart using a genetic algorithm for a multi-objective optimization \cite{maghawry2021approach}}}
    \label{optim_flow}
\end{figure}

The elastic plates are assumed to be aluminum ($\rho = 2,700$ kg/m$^3$, $E=70$ GPa, $\nu =0.334$).  The ribs are considered rigid and of negligible mass, 
and we take $a = d/2$.  A numerical optimization was performed to find the lengths $h_1$, $h_2$, $h_3$ and $d$ for a desired transmission frequency, $f_d$. 
The optimization statement is defined by 
\beq{1=2}
    \text{Cost functions :}
    \begin{cases} & \text{$CF_1$: } \text{Minimize} (- \text{mean} (\text{E}))  ,
    \\
    &\text{$CF_2$: } \text{Minimize} (- $BW$)  ,
   \end{cases}
\eeq
where $\text{E}\le 1$ is the transmitted energy, and $\text{BW}$ is the bandwidth in Hz, with 
\beq{multiobj}
\text{Constraints :}
\begin{cases}
    & 0.5\, \text{mm} \le h_1 \le 3 \, \text{mm},
   \\
   & 4\, \text{mm} \le h_2 \le 24 \, \text{mm},
   \\
   & 0.1 \, \text{mm} \le h_3 \le 1 \, \text{mm},
   \\
   & 2 \, \text{cm} \le d \le 12 \, \text{cm}.
   \\
   & | f_0 - f_d | \le 3 \, \text{Hz}
\end{cases}
\eeq
where $f_0$ is the central peak frequency.  
We consider the desired frequency to be either $f_d=500$ Hz or $f_d=1000$ Hz.
The optimization was implemented in MATLAB using the genetic algorithm \href{https://www.mathworks.com/help/gads/gamultiobj.html}{gamultiobj} suitable for  multi-objective optimization.  The optimal results were obtained by running the program subject to the above optimization statement and   constraints, with Population Size of $400$ and 
Maximum Generation equal to  $100$. {Figure \ref{optim_flow} shows the optimization flowchart we used for this study to achieve optimal results using the genetic algorithm.}

\begin{figure}[H]   
    \centering
    \includegraphics[width=0.7\textwidth]{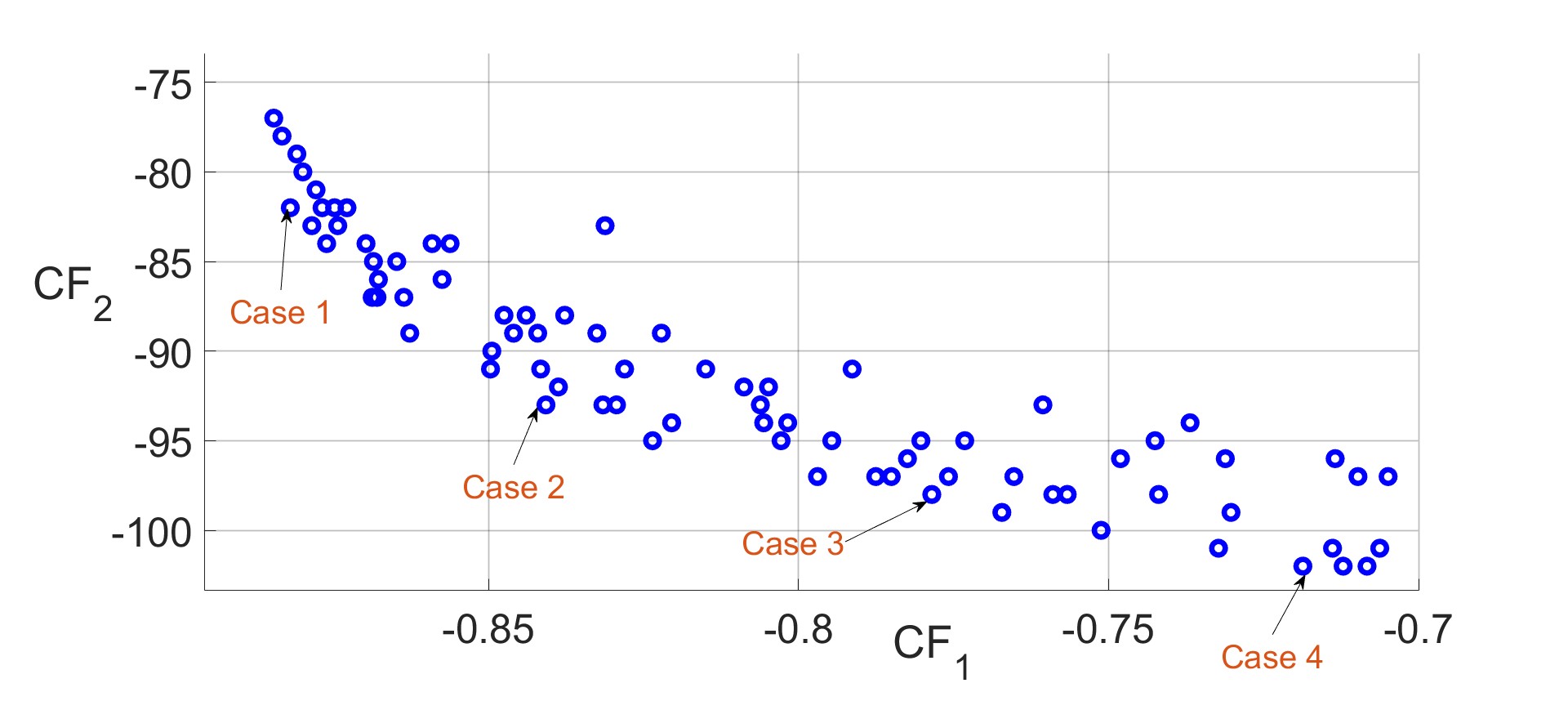}
    \caption{Pareto Front obtained from the optimization 
    \eqref{1=2} and \eqref{multiobj} for $f_0 \approx$  500 Hz.}
    \label{pareto2D_500Hz}
\end{figure}
\begin{figure}[H] 
    \centering
    \includegraphics[width=0.7\textwidth]{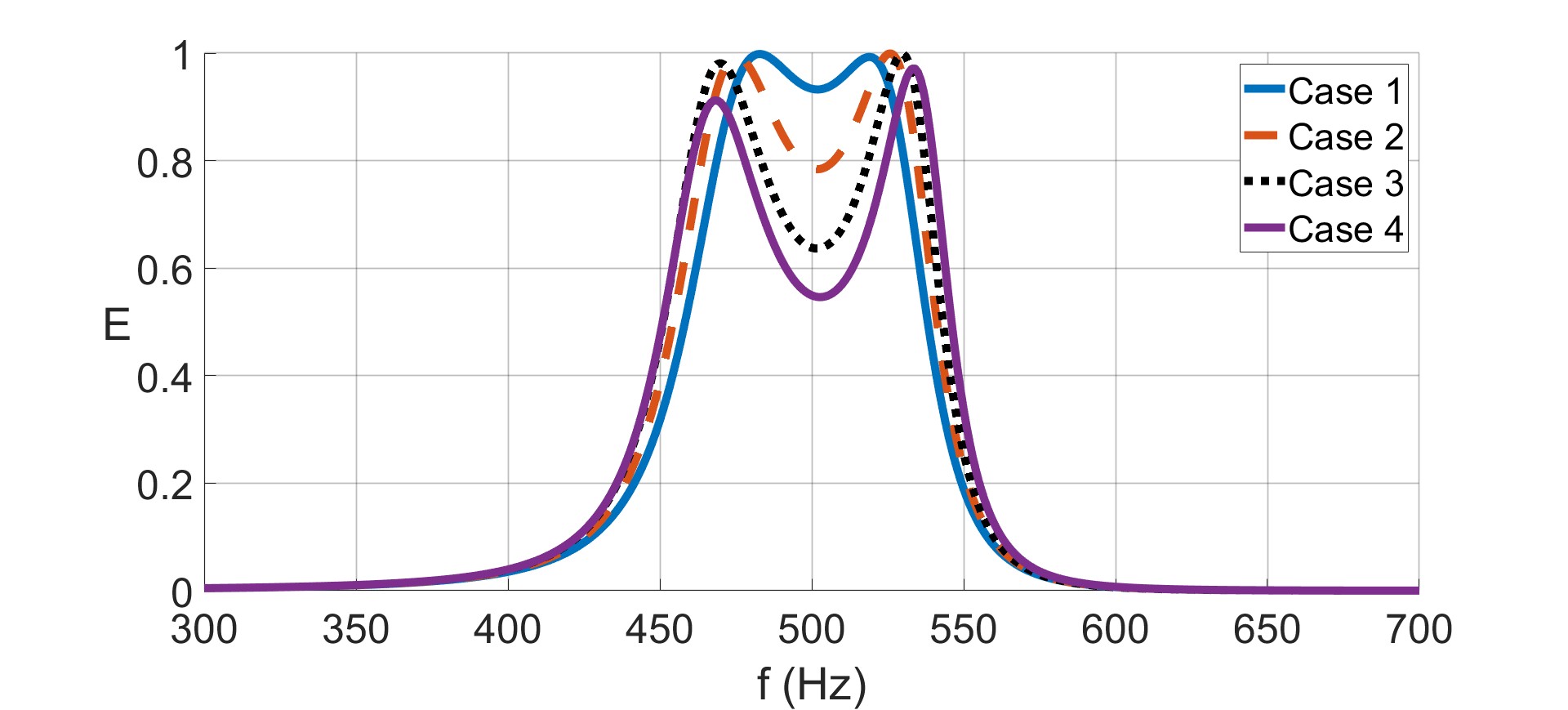}
    \caption{Four optimal cases for   $f_0 \approx 500$ Hz
    with system parameters given in Table \ref{tab1}.}
    \label{4cases_E}
\end{figure}
\begin{figure}[H]
    \centering
    \subfigure[]{\includegraphics[width=0.49\textwidth]{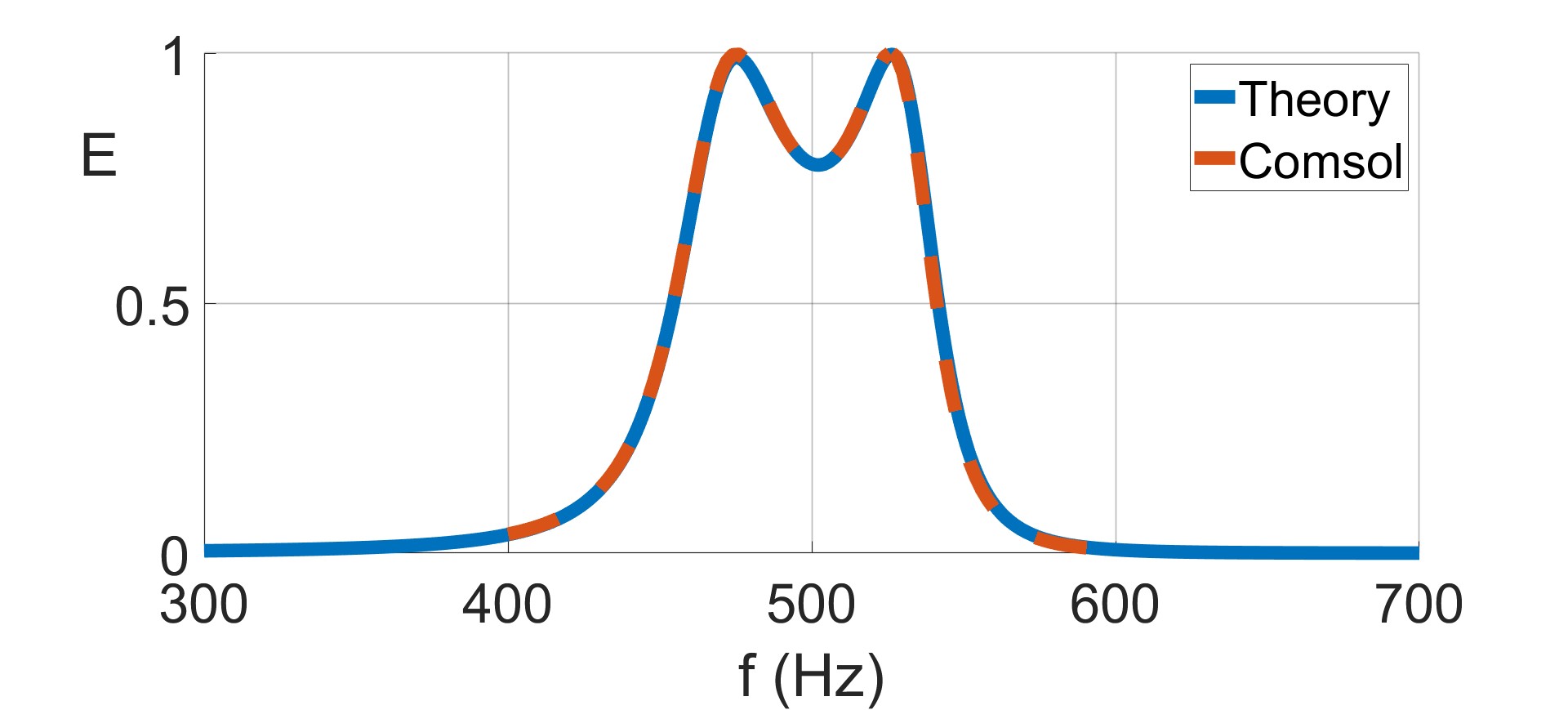}}
    \subfigure[]{\includegraphics[width=0.49\textwidth]{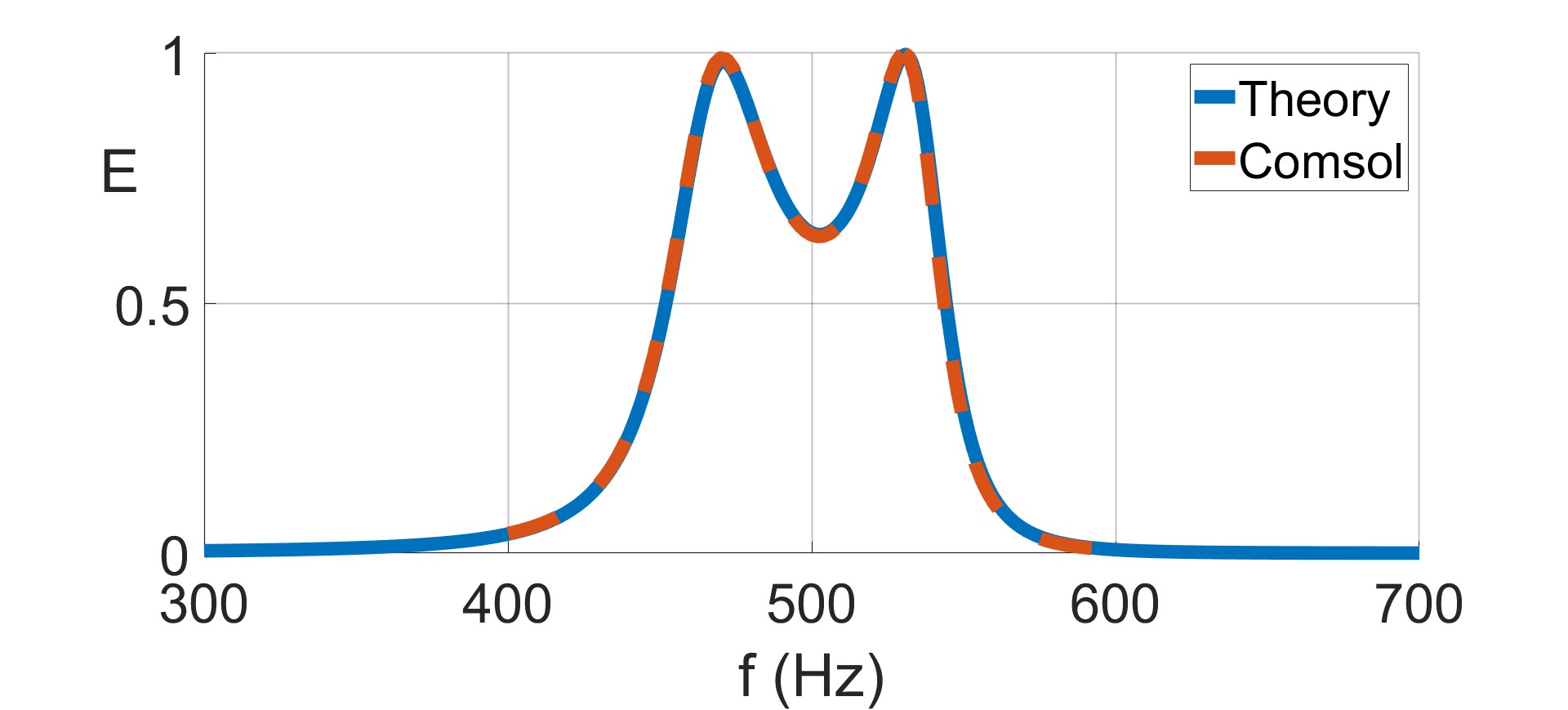}}
    \caption{Verification with Comsol for two cases shown in Fig.\ \ \ref{4cases_E}: (a) Case 2,  (b) Case 3.}
    \label{E_verif}
\end{figure}

The concept of a Pareto Front is useful in characterizing the optimal solutions in a multi-objective optimization.  In this case, we have two objective functions, \eqref{1=2}, for which Fig. \ref{pareto2D_500Hz} shows the Pareto Front obtained for $f_0 \approx$  500 Hz.  
Based on these four cases were selected from the Pareto Front with parameters listed in Table \ref{tab1},  the computed transmitted energy for the four cases is shown in Fig.\  \ref{4cases_E}.  The Matlab simulation is verified by comparison with computational results using Comsol, see Fig.\ 
\ref{E_verif}.

\begin{center} 
\begin{table}
\centering
\begin{tabular}{ |c|c|c|c|c| } 
\hline
Parameters & Case 1 & Case 2 & Case 3 & Case 4 \\
\hline   \hline
$h_1$ (mm) &1.11 &1.34 &1.508 &1.51\\
\hline   \hline
$h_2$ (cm) &1.57 &1.64 &1.61 &1.62\\
\hline   \hline
$h_3$ (mm) &0.335 &0.435 &0.518 &0.532\\
\hline   \hline
$d$ (cm) &6.08 &6.92 &7.56 &7.60\\
\hline
\end{tabular}
\caption {The parameters for the four cases plotted in 
Fig.\ \ref{4cases_E}, selected from the Pareto Front in Fig.\ \ref{pareto2D_500Hz} for $f_0 \approx$  500 Hz.} \label{tab1}
\end{table}
\end{center}

\begin{figure}[H]
    \centering
    \includegraphics[width=0.7\textwidth]{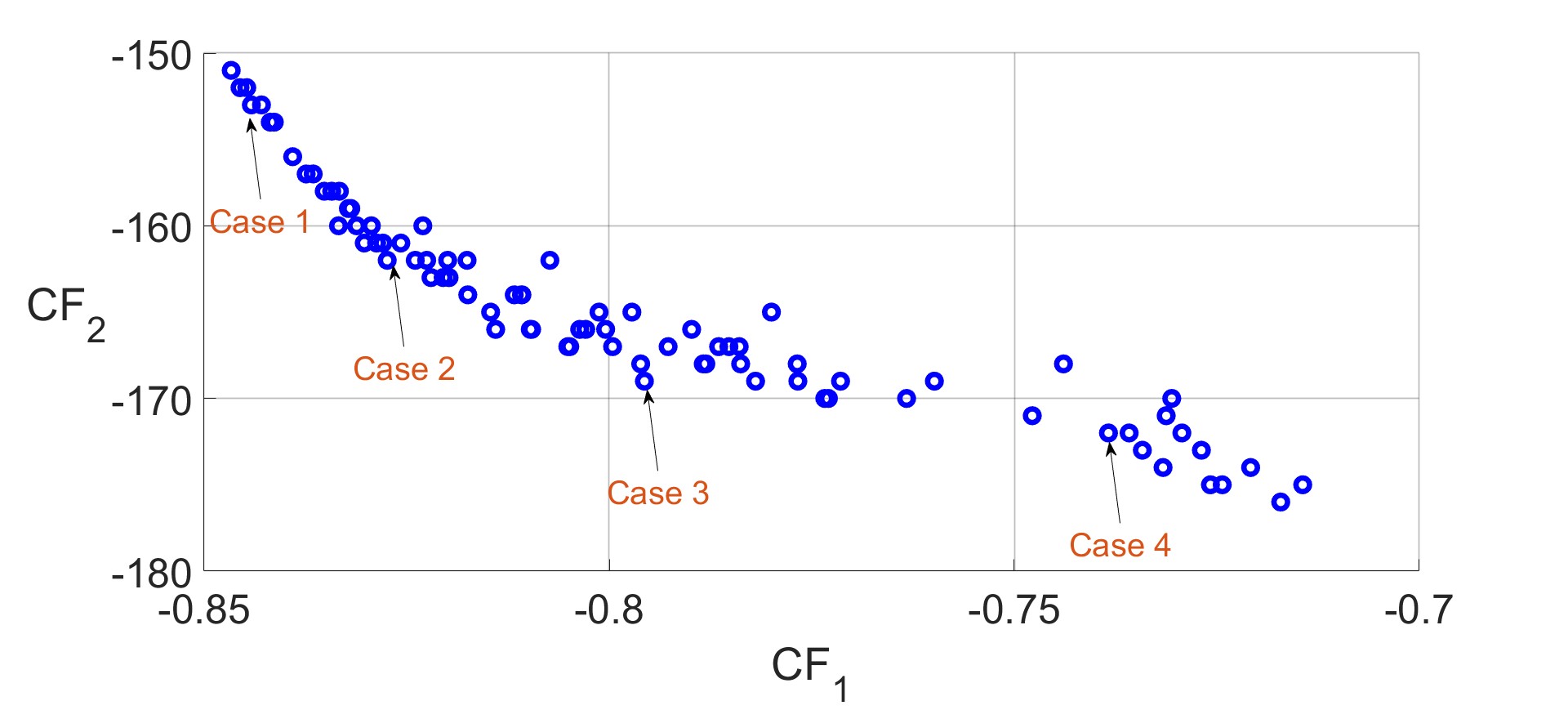}
    \caption{Pareto Front obtained from the optimization 
    \eqref{1=2} and \eqref{multiobj} for $f_0 \approx$  1000 Hz.}
    \label{pareto2D_1000Hz}
\end{figure}

\begin{figure}[H]
    \centering
    \includegraphics[width=0.7\textwidth]{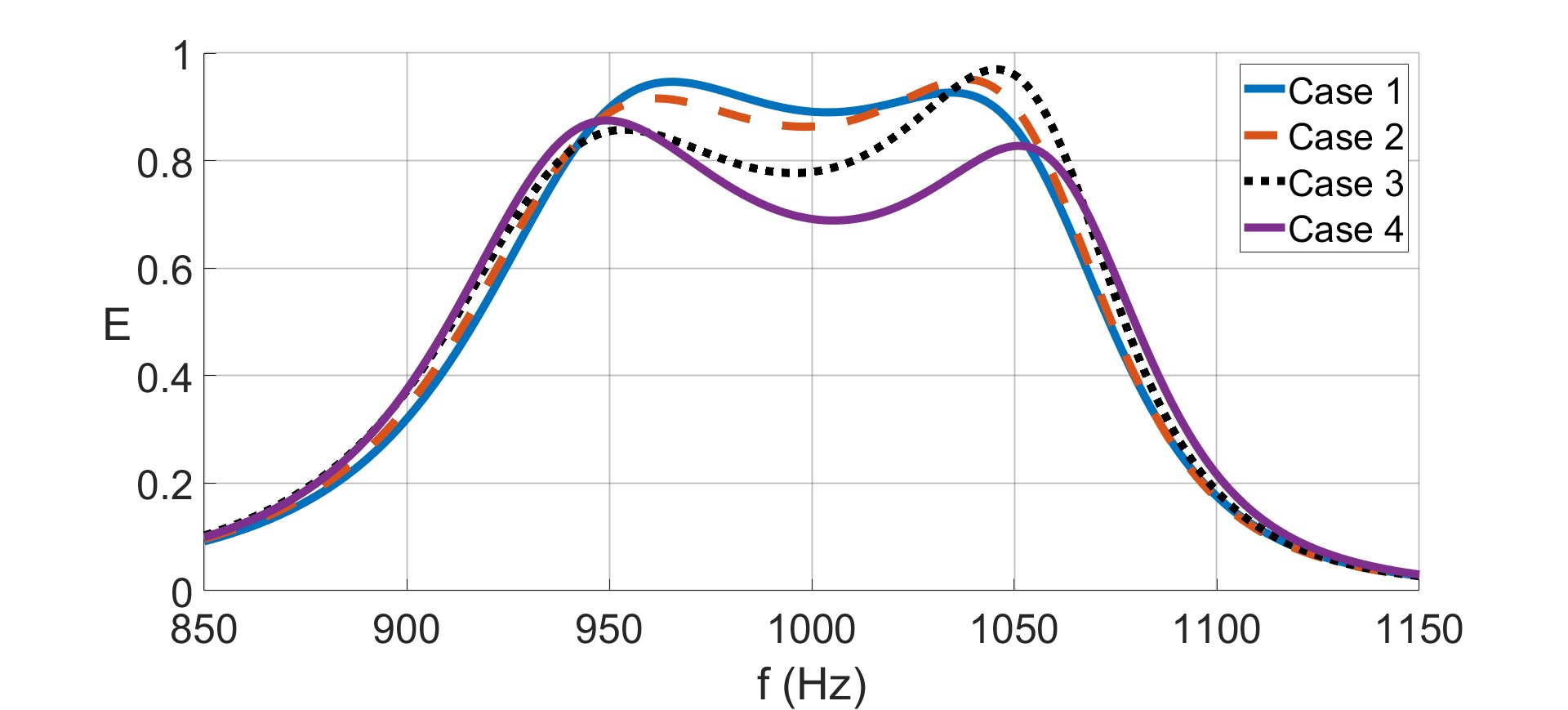}
    \caption{Four optimal cases for   $f_0 \approx 1000$ Hz
    with system parameters given in Table \ref{tab2}.}
    \label{4cases_E_1000Hz}
\end{figure}

Similar results are presented in Figs.\  \ref{pareto2D_1000Hz}
and \ref{4cases_E_1000Hz} for transmission frequency $f_0 \approx$  1000 Hz.

\begin{center}
\begin{table}
\centering
\begin{tabular}{ |c|c|c|c|c| } 
\hline
Parameters & Case 1 & Case 2 & Case 3 & Case 4 \\
\hline  \hline
$h_1$ (mm) &0.911 &0.913 &0.938 &1.08\\
\hline   \hline
$h_2$ (cm) &1.22 &1.20 &1.17 &1.19\\
\hline   \hline
$h_3$ (mm) &0.281 &0.286 &0.304 &0.364\\
\hline   \hline
$d$ (cm) &3.95 &3.98 &4.10 &4.50\\
\hline
\end{tabular}
\caption {The parameters for the four cases plotted in 
Fig.\ \ref{4cases_E_1000Hz}, selected from the Pareto Front in Fig.\ \ref{pareto2D_1000Hz} for $f_0 \approx 1000$ Hz.} \label{tab2}
\end{table}
\end{center}

The most notable features of the transmission curves in Figs.\ 
\ref{4cases_E} and \ref{4cases_E_1000Hz} are the large bandwidths, on the order of 90 Hz and 170 Hz, respectively.  This is particularly significant when compared with the simpler 2-plate model \cite{HBYANN2025}, where bandwidths of approximately 10 Hz and 18 Hz were found.   A common feature of the broadband solutions appears to be that the thickness $h_2$ far exceeds $h_1$ and $h_3$, while $h_1 \approx 3 h_3$; see the summarized results in Tables \ref{tab1} and \ref{tab2}. 

These results and observations are discussed further and explained in the next Sections. 

\section{Simplification  of the optimal system and transmission conditions}\label{sec4} 

\subsection{Optimal transformer as a 2-plate structure with large mass-like impedance}

 \begin{figure}[H]
    \centering
    \includegraphics[width=0.6\textwidth]{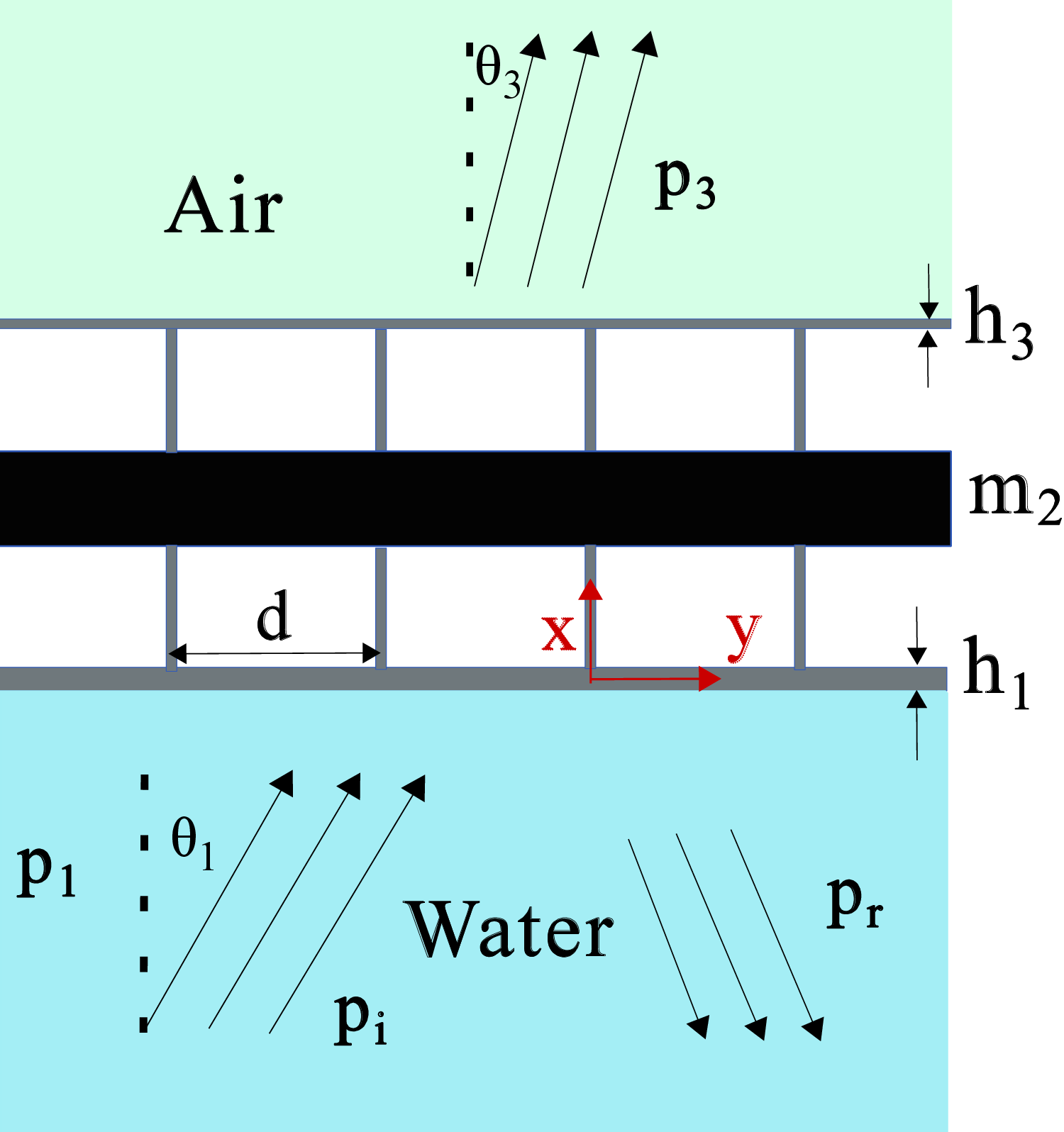}
    \caption{{The   optimized 3-plate system has  a central plate   substantially thicker than the other two plates according to the 
    results  of Tables \ref{tab1} and \ref{tab2}.  The central plate therefore acts as a rigid mass because of its  significantly larger bending stiffness, and the system is  insensitive to the value of the spacing $a$ of Fig.\ \ref{Flex_3Plates}, shown here as $a=0$. Based upon these findings, we simplify the impedance transformer model to a two-plate flex layer with a central mass-like impedance.}    }
    \label{Flex_3Plates_Optimal}
\end{figure}

The numerical results show that in  optimal cases the central plate is far thicker than the others, and  therefore, it acts as a translating mass  with little flexural bending, Fig.\ \ref{Flex_3Plates_Optimal}.  From its definition in Eq.\ \eqref{X} it follows, using   the large bending stiffness,  that  
\beq{783}
S_{p2}^{(\phi)} (0) \approx  \hat{Y}_{p2} (0) = 
\frac 1{-\ii \omega m_2}  \equiv s_{p2} 
\eeq
which is an effective mass.    This simplification implies that the optimal transformer can be considered as a {\em two-plate system with  a mass-like internal  impedance}, as we show next. 

Using the results from Eq.\ \eqref{7073}, the reflection coefficient for normal incidence becomes
\beq{70736}
   R(0) \approx  R_1(0)\,  { \big( s_{p2} + S_3 (0) \big) }{ b^{-1}(0)} \, 
   \Gamma_1 (0)
\eeq
where 
\beq{7-36}
   \Gamma_1 (0 )=   
    \frac   { b(0)} {  s_{p2} + S_3 (0)  } 
   -  \hat Y_{1}(0) 
    +  \frac 1{\hat Z_{p1}(0) - \hat Z_{f1}(0)} 
\eeq
and  
\beq{89} b (\xi) = S_1 (\xi)S_3 (\xi) + \big(  S_1 (\xi)+S_3 (\xi)\big)  s_{p2} .
\eeq

Consider a two-plate flex-layer with plates  1 and 3 separated by  impedances $Z_{0\pm}$, and then let $1/ Z_{0-} \to 0$, then we obtain exactly the same result as above if \cite{HBYANN2025} $s_{p2} = \frac d{4Z_{0+} }$, i.e. 
\beq{147}
Z_{0+} = - \frac{ \ii}4  \omega m_2 d . 
\eeq
This value for $Z_{0+}$ is consistent with the model of an impedance as a spring-mass system of mass $m_2 d$ with a very stiff spring 
\cite[Eq.\ (A7)]{HBYANN2025}.

Hence, the optimal three-plate system reduces to a two-plate flex layer with a central mass defined by impedance $Z_{0+}$ of \eqref{147}.   The search for full transmission then reduces to finding the impedance (or central mass) and the properties of the two plates facing air and water. 

\subsection{Asymptotic approximations and  alternative transmission conditions}

The two conditions in Eq.\ \eqref{330} are equivalent to the full transmission condition 
 $ \Gamma_1 (0) =0$  (or  $ \Gamma_3 (0) =0$), as has been tested for the examples above.  
  Based on the results of Tables \ref{tab1} and \ref{tab2}, and the fact that {$Z_1$ and $Z_3$ are 1.5 MRayl and 408 Rayl, respectively,  in all of the examples considered we have 
 $Z_1^2 \gg (\omega_0 m_1)^2$ and 
  $Z_3^2 \ll (\omega_0 m_3)^2$; in fact max $(\omega_0 m_1/Z_1)^2 =1.5\, 10^{-4}$ and  max $Z_3^2 / (\omega_0 m_3)^2 =0.02$. }   
  The condition \eqref{555}
  can therefore be replaced by the simpler 
  \beq{1-42}
\left| \frac{ S_1 (0) +S_{p2}^{(0)} (0) } {  S_3 (0) + S_{p2}^{(0)} (0) }\right|^{1/2}
  \approx \frac{ \omega_0 m_3 } {Z_e}  \quad\text{where}\quad Z_e \equiv \sqrt{Z_1 Z_3}.
\eeq
 The left hand member of \eqref{1-42} is $\alpha_3$ of Eq.\ \eqref{5+3}. The condition  \eqref{1-42} can also be justified as an asymptotic 
  approximation by expressing \eqref{555} using the small parameter $\epsilon = Z_3/Z_1$ and ignoring terms of O$(\epsilon )$.   For the air/water system with 
$\rho_a = 1.2$ kg/m$^3$, 
$c_a = 340$ m/s, 
$\rho_w = 1000$ kg/m$^3$, 
$c_w = 1500$ m/s, 
  { and hence our small parameter is 
\beq{0933}
  \epsilon = \frac{Z_3}{Z_1}
  = \frac{Z_\text{air} }{Z_\text{water}}
  = 2.72 \, \times 10^{-4}.
  \eeq }

Finally, { the observation from Eq.\ \eqref{783} that optimal systems are essentially two-plate flex layers with a central mass,  combined with the asymptotic approximation 
\eqref{1-42}, allow us to replace 
 the  two conditions in Eq.\ \eqref{330}   by  simpler ones, }
\beq{444}
\begin{aligned}  \big|  1-\ii \omega_0  m_2  S_1 (0) \big|
 & \approx \frac{\omega_0 m_3}{ Z_e},
\\
\big| 1-\ii \omega_0  m_2  S_3 (0) \big| 
 & \approx \frac{ Z_e}{\omega_0 m_3} . 
\end{aligned} 
\eeq

\subsection{A unique class of solutions to the  transmission conditions}

We consider a particular type of solution to the transmission conditions \eqref{444}.  Specifically, we seek solutions to the  pair of complex-valued equations
\beq{44-}
\boxed{ \begin{aligned} 
 1-\ii \omega_0  m_2  S_1 (0)
 & \approx -\ii \frac{\omega_0 m_3}{ Z_e},
\\
  1-\ii \omega_0  m_2  S_3 (0)  & \approx -\ii \frac{ Z_e}{\omega_0 m_3} . 
\end{aligned} }
\eeq
It is clear that all solutions of \eqref{44-} are solutions of \eqref{444}.   
The motivation behind  \eqref{44-} is twofold: first we observed that solutions to the complex pair of equations 
 agree with multiple  numerical results for the optimized systems.  A second and more physical reason is discussed in the next Section in terms of pairs of resonances. For now we explore the analytical consequences of  \eqref{44-} in terms of explicit solutions. 

The quantities on the left in \eqref{44-} are,  for $j=1,3$, 
\bal{3--4}
    1-\ii \omega_0  m_2  S_j (0) &=
 1 + \frac{m_jm_2 \omega_0^2}{Z_j^2 +(m_j \omega_0)^2 }
 \notag \\ 
 &-\sum_{n\ne 0} \frac { m_2 \omega_0^2 }
 {D_j\xi_n^4 - \big(m_j + \frac{\rho_j}{\sqrt{\xi_n^2-k_j^2}} \big)\omega_0^2}
 - \frac{\ii m_2 \omega_0 Z_j}{Z_j^2 +(m_j \omega_0)^2 } .
 \eal
It may be checked that the appropriate approximations are 
 \beq{6-34}
 \begin{aligned} 
1-\ii \omega_0  m_2  S_1 (0) &\approx 
1-  \frac{m_2 \omega_0^2 d^4}{720 D_1}
-  \ii \frac{m_2 \omega_0  }{Z_1} , 
\\
1-\ii \omega_0  m_2  S_3 (0)  &\approx 
1 + \frac{m_2}{m_3}
   \frac {\gamma}4 \big( \cot  \frac {\gamma}2  +\coth  \frac {\gamma}2
   \big)
-  \ii \frac{m_2 Z_3}{m_3^2 \omega_0}
 \end{aligned} 
 \eeq
 where $\gamma =\big( m_3 \omega_0^2 /D_3\big)^{1/4} d$
 (see Eqs.\  \eqref{45} and \eqref{452}).
  Comparing Eqs.\ \eqref{44-} and \eqref{6-34} implies that the 
  two equations obtained by equating the imaginary parts in the former reduce to a single relation.  Then setting the real parts of   \eqref{44-} to zero yields two additional equations.   In summary, we obtain three relations: 
\bse{554}
\bal{660}
 m_3 &\approx \epsilon^{1/2}\, m_2, 
 \\
 m_2 \omega_0^2 d^4  &\approx 720 D_1,
  \label{756}
 \\
 m_3 \omega_0^2 d^4 \ &\approx 500 D_3 , 
 \label{757}
\eal
\ese
The last result uses the fact, based on \eqref{660},
that the zero of $\Re \big( 1-\ii \omega_0  m_2  S_3 (0) \big)$  is  $\gamma \approx 4.73 + 0.8164\, \epsilon^{1/2} $.

Combining \eqref{660}, \eqref{756} and  \eqref{757} implies, assuming the same material in plates 1 and 3, that 
$h_3 \approx  1.129\,\epsilon^{1/6}\, h_1 $
which for air/water translates to 
$ h_3 \approx  0.287\, h_1 $. 
If all plates have the same density the relative thicknesses are, in terms of the thickest, plate 2, 
\beq{446}
\begin{aligned}
h_1 &\approx 0.886\,\epsilon^{1/3}\, h_2,
\\
h_3 &\approx   \epsilon^{1/2}\, h_2,
\end{aligned}
\eeq
which means for air/water that 
$h_1\approx 0.057\, h_2$  and 
$h_3 \approx  0.016\, h_2$  (and $h_1\approx 3.49 \, h_3$).  Whether or not the materials in the plates are the same, Eqs.\ \eqref{554} imply that {\em the relations between the plate thickness are independent of transmission frequency}.  Selecting a value for one of the three thickness then defines the other two through the asymptotic parameter $\epsilon$.   The remaining system dimension, the rib spacing, follows from either \eqref{756} or \eqref{757} as a function of frequency according to   $d \propto \omega_0^{-1/2}$.  It is interesting to note from \eqref{446} a relation between the plate thicknesses that is independent of the impedance ratio: $1.44\, h_1^3 \approx h_2h_3^2$. A modified version of this approximate identity is presented in Section \ref{sec5}. 

In summary, the three explicit relations \eqref{554} follow from the complex-valued transmission conditions \eqref{44-}.    
Alternate derivations of \eqref{756} and \eqref{757} are  presented in the next Section where we interpret the optimized  solution in terms of system resonances.

\subsection{Numerical verification of the asymptotic solutions}

Examples are presented to test the accuracy of the asymptotic approximations discussed above, particularly Eqs.\ \eqref{554} and \eqref{446}.
 We start by choosing the thickness $h_2$ of the central mass-like plate,  assuming the three plate are all the same material, aluminum.   The two equations \eqref{446} imply approximated plate thicknesses $\tilde h_1$ and $\tilde h_3 $. We then choose the target transmission frequency,  $f_0$, and use either \eqref{756} or \eqref{757} to find the associated optimal value for the approximated rib spacing $\tilde{d}$. 
More accurate results for $h_1$, $h_3$, and $d$  are found for the chosen  $h_2$ and $f_0$ by  solving Eqs.\ \eqref{444} and \eqref{44-} numerically, using the MATLAB function \href{https://www.mathworks.com/help/optim/ug/fsolve.html}{fsolve}. 

\begin{figure}[h]
    \centering
    \includegraphics[width=0.65\textwidth]{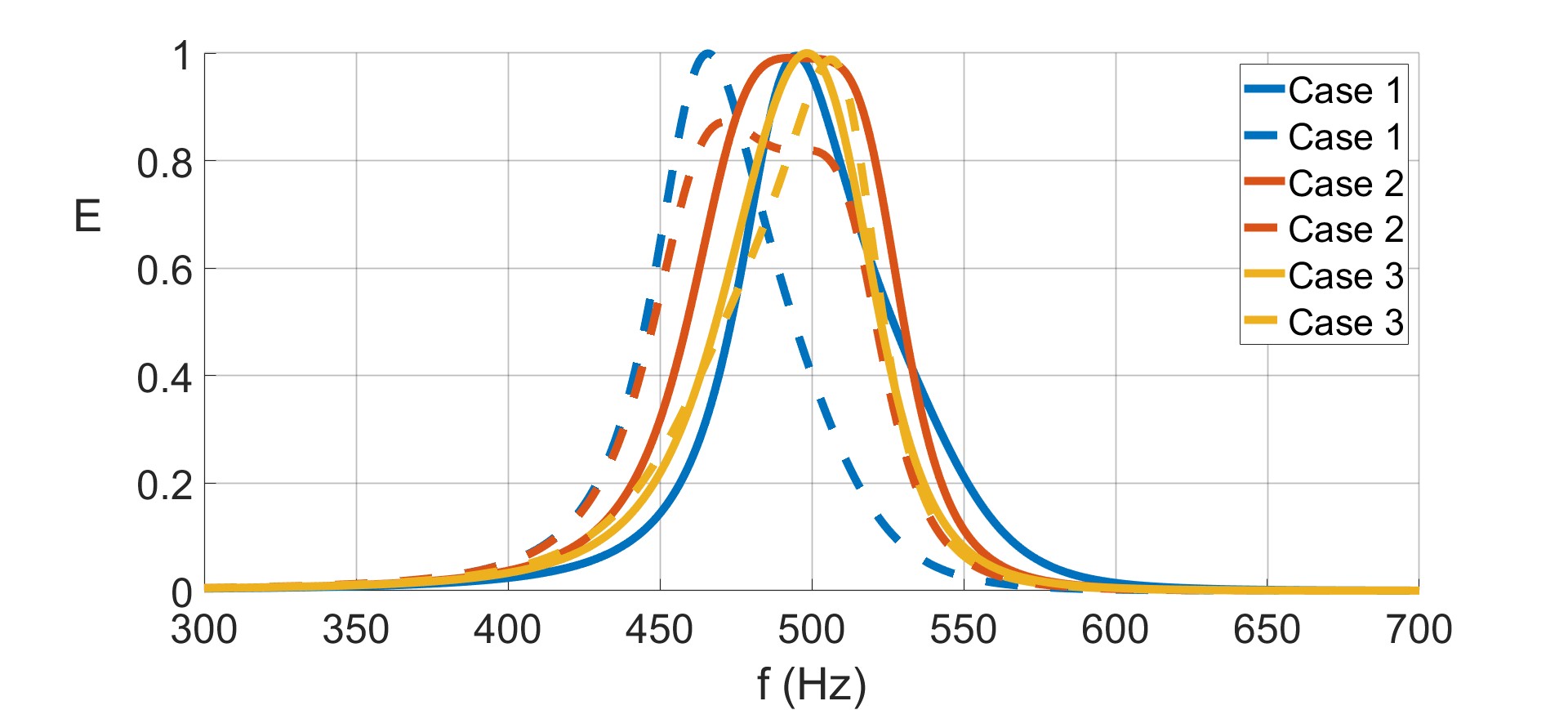}
    \caption{Transmittivity at $f_0 = $ 500 Hz for Case 1 ($h_2 =$ 1 cm),  Case 2 ($h_2 =$ 2 cm), and Case 3 ($h_2 =$ 4 cm). The dashed lines are the results from approximated values $\tilde h_1$, $\tilde h_3$, and $\tilde{d}$, see Table \ref{tab3_h1h2h3d}.  }
    \label{f0_500Hz_h1_h2_h3_d}
\end{figure}

\begin{figure}[H]
    \centering
    \includegraphics[width=0.7\textwidth]{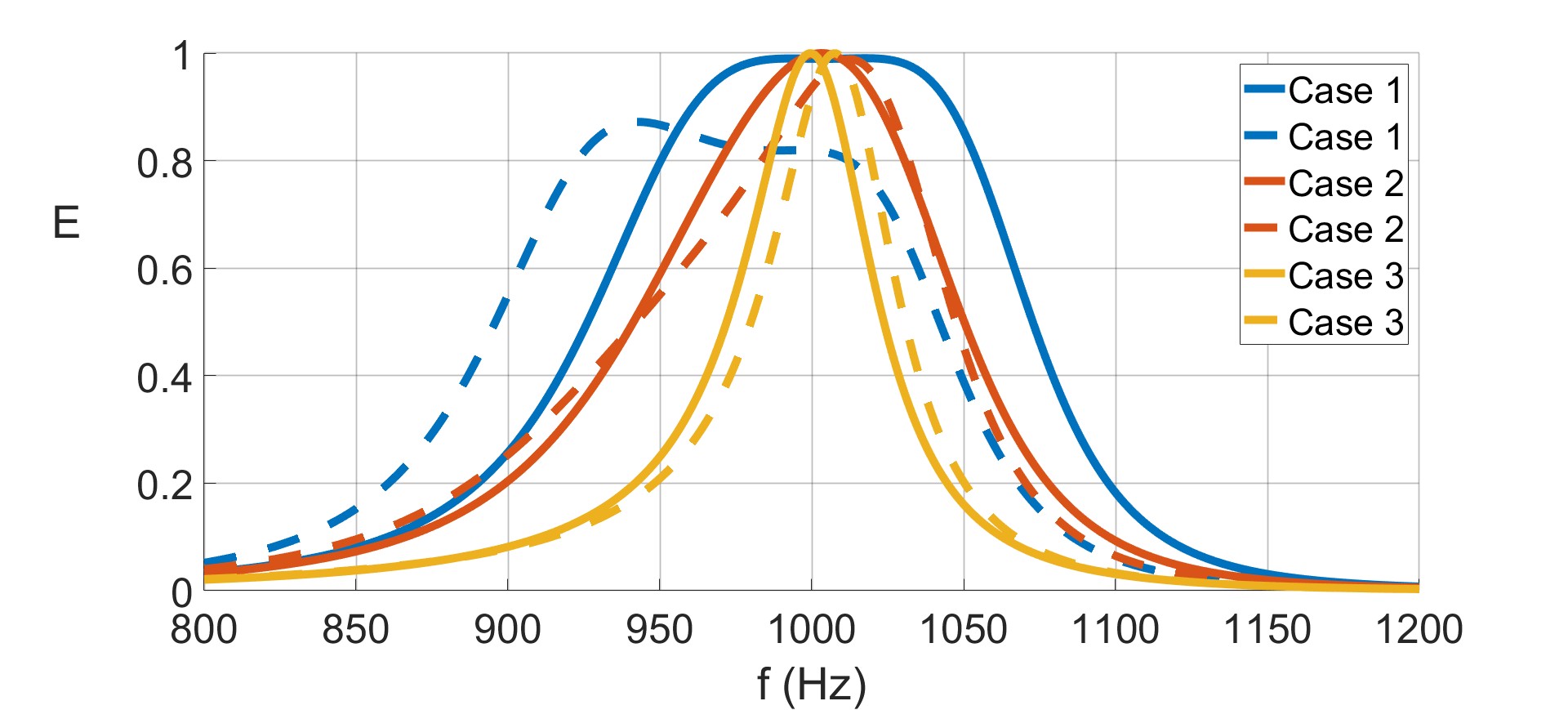}
    \caption{Transmittivity at $f_0 = $ 1000 Hz for $h_2 =$ 1 cm,  2 cm and 4 cm, see Table \ref{tab4_h1h2h3d}. The dashed lines are the results from approximated values $\tilde h_1$, $\tilde h_3$, and $\tilde{d}$.  }
    \label{f0_1000Hz_h1_h2_h3_d}
\end{figure}

Three  values are taken for the central plate thickness: $h_2= $ 1 cm, 2 cm and 4 cm, and two target frequencies are chosen: $f_0 = $ 500 Hz and  1000 Hz. 
Table \ref{tab3_h1h2h3d}  shows the approximate and improved values for the thickness of plates 1 and 3 and the rib separation length  for each of the three values of $h_2$ with  $f_0 = $ 500 Hz.  Energy transmission for the approximate and refined models are shown 
in Fig.\ \eqref{f0_500Hz_h1_h2_h3_d}. 
The corresponding results for $f_0 = $ 1000 Hz are given in Table \ref{tab4_h1h2h3d}
and  Fig.\ \eqref{f0_1000Hz_h1_h2_h3_d}.

\begin{center} 
\begin{table}
\centering
\begin{tabular}{ |c|c|c|c| } 
\hline
Parameters & Case 1 & Case 2 & Case 3 \\
\hline   \hline
$h_2$ (cm)  & 1.0 & 2.0 & 4.0 \\
\hline   \hline
\multirow{2}{4em}{$\tilde{h}_1$ (mm)\\${h_1}$ (mm)} 
&0.570 &1.140 &2.280\\
&0.567 &1.183 &2.346\\
\hline   \hline
\multirow{2}{4em}{$\tilde{h}_3$ (mm)\\${h_3}$ (mm)}
&0.163 &0.327 &0.653\\
&0.163 &0.327 &0.653\\
\hline     \hline
\multirow{2}{4em}{$\tilde d$ (cm)\\${d}$ (cm)} 
&4.25 &6.02 &8.51\\
&4.11 &6.04 &8.54\\
\hline
\end{tabular}
\caption {The approximate ($\bar h_1$ etc. obtained from Eqs.\ \eqref{446} and \eqref{756}) and refined values of $h_1$, $h_3$ and $d$ for the three chosen values of the central plate thickness $h_2$ and $f_0=$ 500 Hz.  The associated transmittivities are  plotted in 
Fig.\ \ref{f0_500Hz_h1_h2_h3_d}.} \label{tab3_h1h2h3d}
\end{table}
\end{center}

\begin{center} 
\begin{table}
\centering
\begin{tabular}{ |c|c|c|c| } 
\hline
Parameters & Case 1 & Case 2 & Case 3 \\
\hline   \hline
$h_2$ (cm)  & 1.0 & 2.0 & 4.0 \\
\hline   \hline
\multirow{2}{4em}{$\tilde{h}_1$ (mm)\\${h_1}$ (mm)} 
&0.570 &1.140 &2.280\\
&0.591 &1.173 &2.326\\
\hline   \hline
\multirow{2}{4em}{$\tilde{h}_3$ (mm)\\${h_3}$ (mm)}
&0.163 &0.327 &0.653\\
&0.163 &0.327 &0.653\\
\hline   \hline
\multirow{2}{4em}{$\tilde d$ (cm)\\${d}$ (cm)} 
&3.01 &4.26 &6.02\\
&3.02 &4.27 &6.03\\
\hline
\end{tabular}
\caption {The approximate ($\bar h_1$ etc. obtained from Eq.\ \eqref{446} and \eqref{756}) and refined values of $h_1$, $h_3$ and $d$ for the three chosen values of the central plate thickness $h_2$ and $f_0=$ 1000 Hz.  The associated transmittivities are  plotted in Fig.\ \ref{f0_1000Hz_h1_h2_h3_d}.} \label{tab4_h1h2h3d}
\end{table}
\end{center}

We conclude from the examples presented that for given values of $h_2$ and the transmission frequency $f_0$ the asymptotic results in  Eqs.\ \eqref{554} and \eqref{446} provide very accurate approximations to the system parameters $h_1$, $d$, and especially  $h_3$. 

\begin{figure}[H]
    \centering
    {\includegraphics[width=0.7\textwidth]{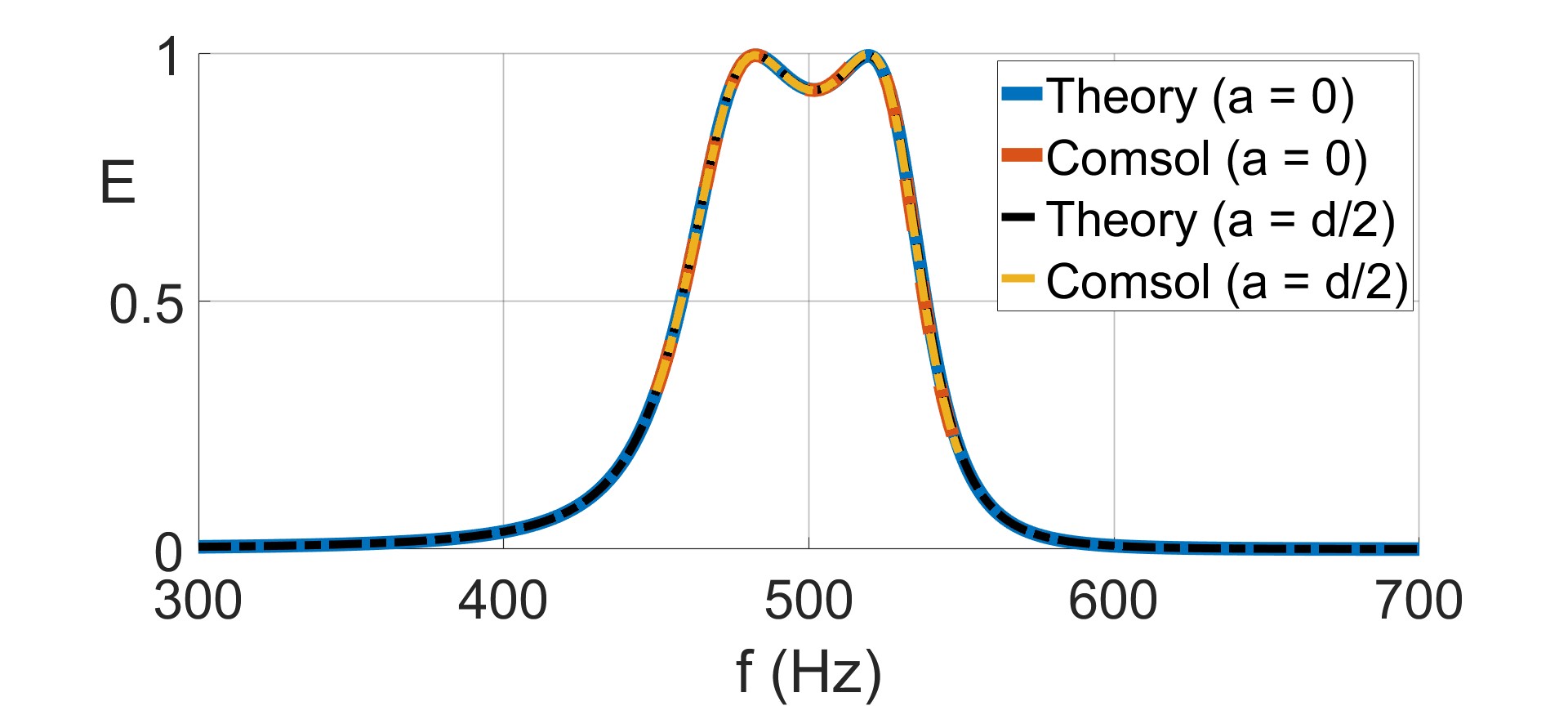}}
    \caption{ The  effect of the spacing $a$,  see Fig.\   \ref{Flex_3Plates}, has no effect on the optimized design with the heavy central mass.  Two sets of curves are plotted for $a = 0$ and  $a = d/2$.  In each case theory and Comsol are used, for a total of four identical curves. ($h_1 = 1.11 $ mm, $h_2 = 1.57$ cm, $h_3 = 0.335 $  mm, $d = 6.08$ cm}
    \label{123aeffect}
\end{figure}

Finally, since the optimized solutions all have the thickness of the central plate 2 much greater than the other thicknesses it follows that the second plate bending will be negligible; as a result,  the parameter $a$ 
(see Fig.\ \ref{Flex_3Plates}) should have a very small effect on the displacement of the central plate.  The optimal design of Fig.\ \ref{Flex_3Plates_Optimal} assumes that the spacing parameter $a=0$, see Fig.\ \ref{Flex_3Plates}.
As seen in Fig.\ (\ref{123aeffect}), the value of the parameter $a$ will not change the energy transmission.

\subsection{Addressing the issue of thin thicknesses}\label{4.5}

\begin{figure}[h]
    \centering
    \includegraphics[width=0.5\textwidth]{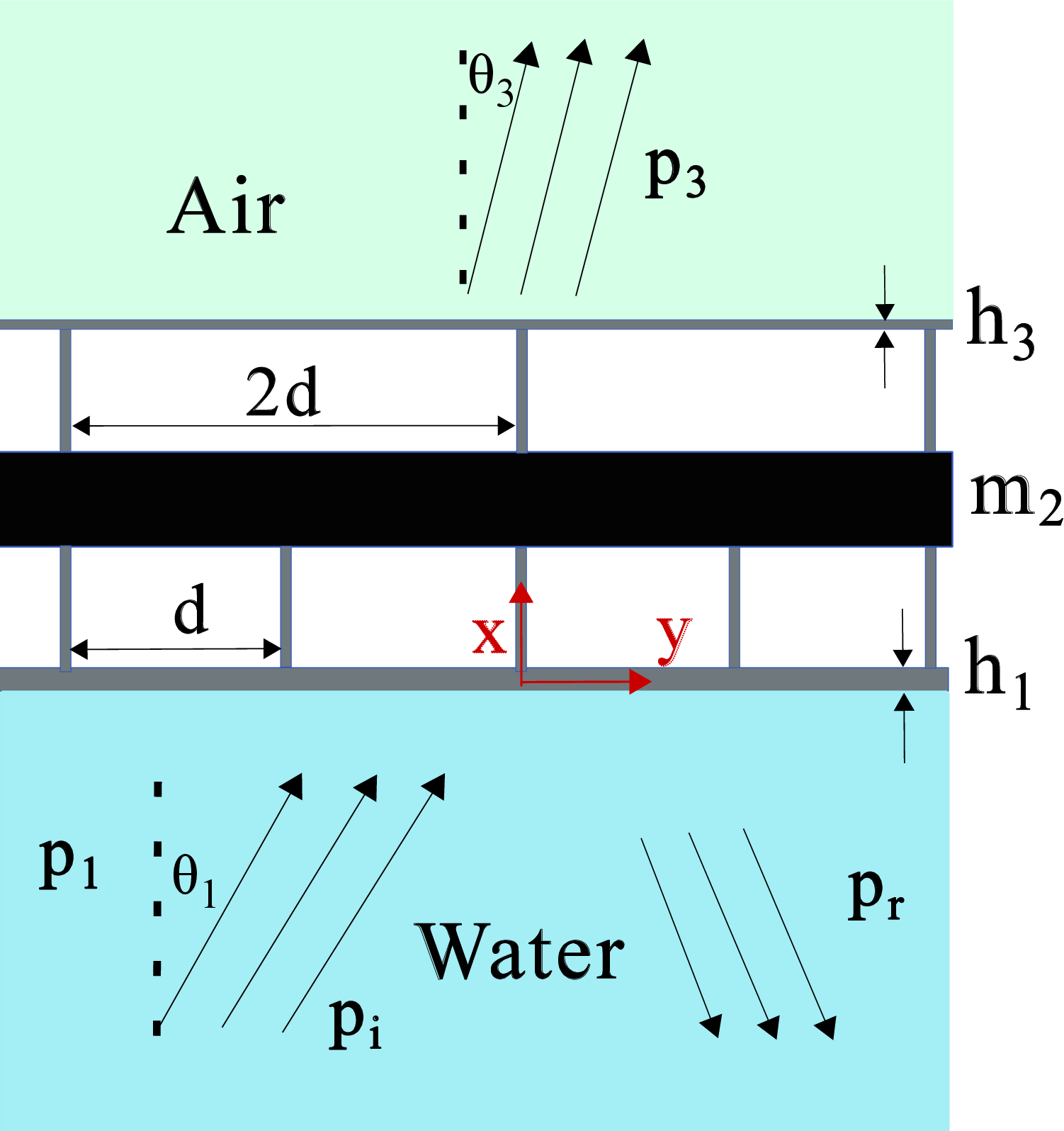}
    \caption{ The new structure is a two-plate flex layer  of aluminum with a central mass-like impedance made of Steel, with different {rib spacings on the two sides} of the flex layer. }
    \label{Flex_3Plates_Optimal_2d}
\end{figure}
{The air-side plate thickness $h_3$ is quite thin   in the examples  of the previous Section, see Tables \ref{tab3_h1h2h3d} and \ref{tab4_h1h2h3d}. In order to increase $h_3$ without significantly changing the entire structure, we modified the transformer to allow for  different rib spacings on the two sides of the flex layer. 
Doubling the spacing on the air side, it follows from  Eq.\ \eqref{757} that the thickness of  plate 3 must increase by a factor of four  in order to maintain the equivalent stiffness. 
In addition, in order to keep the thickness of the central plate in the same range as before, the  plate is considered to be Steel.} The new  flex layer design is depicted in Fig.\ \ref{Flex_3Plates_Optimal_2d}.

\begin{figure}[H]
    \centering
    \includegraphics[width=0.6\textwidth]{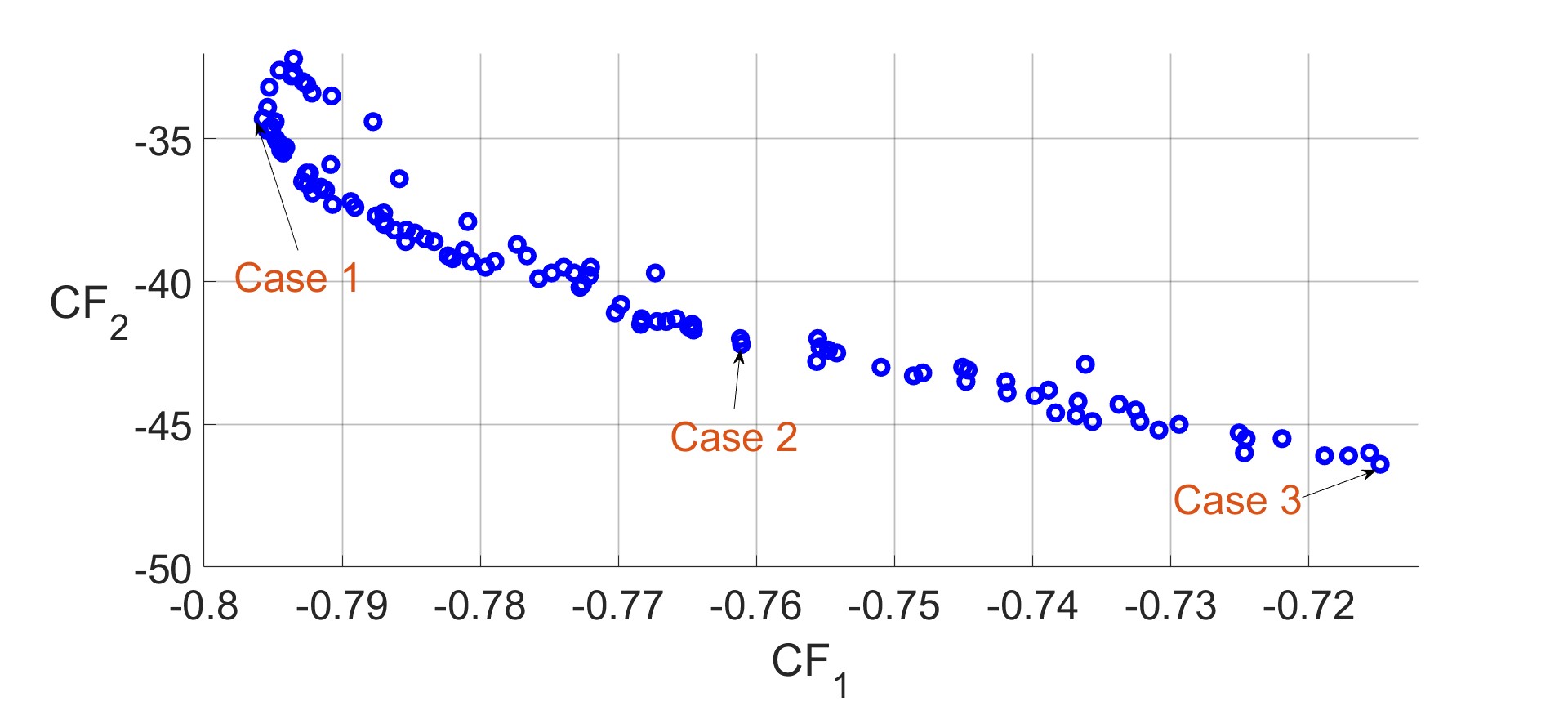}
    \caption{Pareto Front obtained from the optimization 
    at $f_0 \approx$  500 Hz for the modified structure of Fig.\ \ref{Flex_3Plates_Optimal_2d}, }
    \label{Pareto_3Plates_Optimal_2d}
\end{figure}

\begin{figure}[H]
    \centering
    \includegraphics[width=0.7\textwidth]{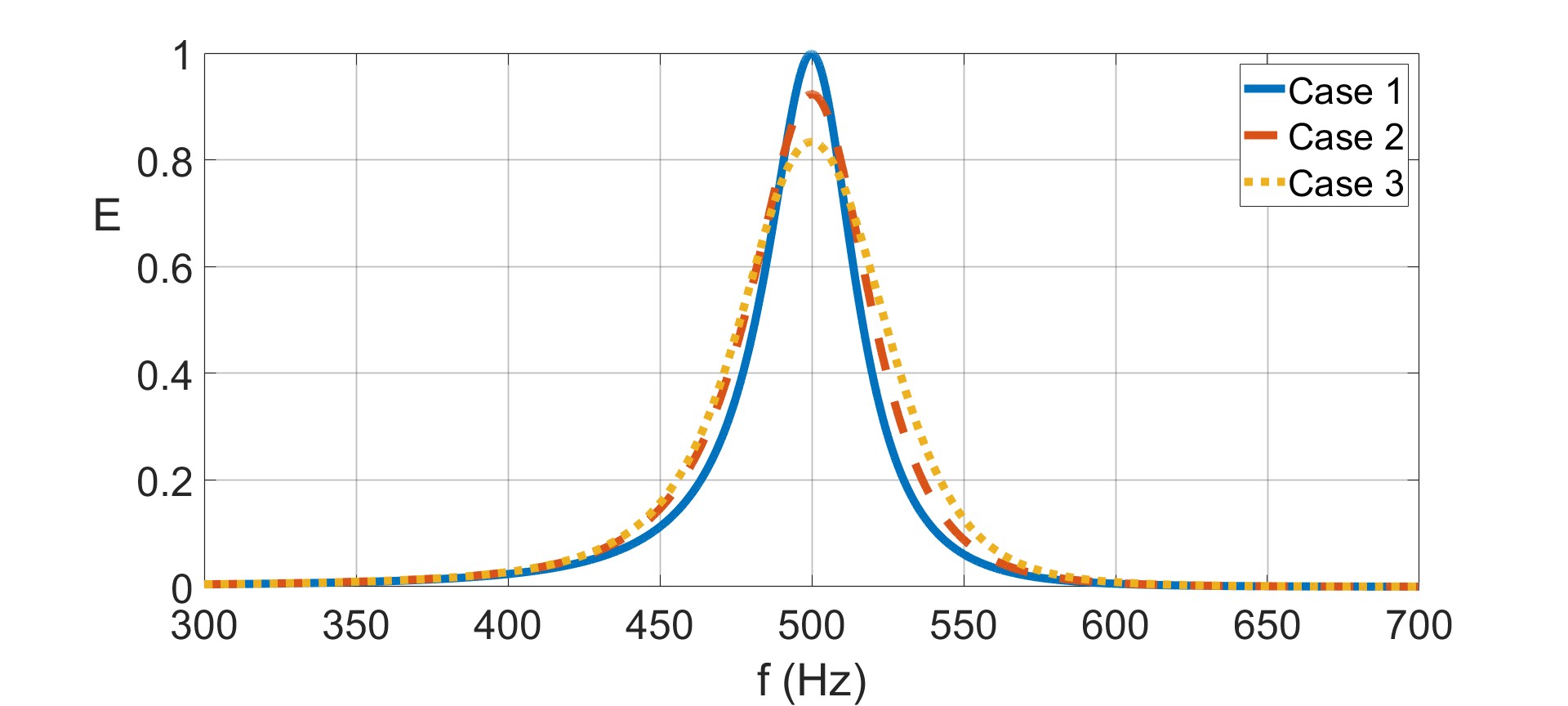}
    \caption{Energy transmission at  $f_0 \approx 500$ Hz for 
     the modified structure of  Fig.\ \ref{Flex_3Plates_Optimal_2d}, with parameters  given in Table \ref{table_2dSpacing}.}
    \label{Pareto_3Plates_Optimal_2d_cases}
\end{figure}

We ran the optimization for {the new structure presented in Fig.\ \ref{Flex_3Plates_Optimal_2d},  resulting in the Pareto Front  illustrated in Fig.\ \ref{Pareto_3Plates_Optimal_2d}. The  parameters for the three cases selected from Fig.\ \ref{Pareto_3Plates_Optimal_2d} are listed in Table \ref{table_2dSpacing}. These results show that $h_3$  is increased more than $h_2$, when compared with Tables \ref{tab3_h1h2h3d} and \ref{tab4_h1h2h3d}.  Energy transmission for the three cases selected from the Pareto Front are shown in Fig.\ \ref{Pareto_3Plates_Optimal_2d_cases}. The bandwidth is  in the same  range as before,  and the issue of very thin thickness $h_3$ is resolved.}

\begin{center}
\begin{table}
\centering
\begin{tabular}{ |c|c|c|c| } 
\hline
Parameters & Case 1 & Case 2 & Case 3 \\
\hline  \hline
$h_1$ (mm) &1.445 &1.433 &1.435\\
\hline   \hline
$h_2$ (cm) &2.08 &1.74 &1.56\\
\hline   \hline
$h_3$ (mm) &1.029 &1.110 &1.168 \\
\hline   \hline
$d$ (cm) &5.349 &5.554 &5.688\\
\hline
\end{tabular}
\caption {The parameters for the three cases   
selected from the Pareto Front in Fig.\ \ref{Pareto_3Plates_Optimal_2d} for the modified flex-layer of Fig.\ \ref{Flex_3Plates_Optimal_2d}.} \label{table_2dSpacing}
\end{table}
\end{center}

\section{An Equivalent Two Degree of Freedom System }\label{sec5} 

{In the previous section, we found simple relations between the system dimensions ($h_1$, $h_2$, $h_3$ and $d$) based on asymptotic approximations of the exact solution.  Here we provide physical/mechanical explanations for these results, in the process finding new relations between the system parameters.  Our starting point is the result from a previous paper \cite{HBYANN2025} for the simpler two-plate flex layer impedance transformer  that the model is 
 equivalent to a single degree of freedom  mass-spring system with resonant frequency equal to the transmission frequency.} 
 The  analogy  helps us understand the underlying mechanical principles operating in the more complicated system. {Thus, we propose} that the current model is analogous to a two degree of freedom (2DOF) system  represented as $ \, \, \, \, \, \hlin  \overset{\kappa_1}{\spring}  {\boxed{\mu_1}}  \,\,\,
\hlin  \overset{\kappa_2}{\spring}  {\boxed{\mu_2}}  \,  $,
where $\kappa_1$, $\kappa_2$ are springs, and ${\mu_1}$, ${\mu_2}$ are masses. This lumped parameter model is discussed in  \ref{appB}.   

{Our objective } is to relate parameters of the 3-plate flex-layer to those of the 2DOF system.  In particular, we identify $\kappa_1$  and   ${\mu_1}$ with the bending stiffness of  plate 1 on the water side and the mass of the central plate, respectively.  The second spring-mass pair, $\kappa_2$,    ${\mu_2}$, will be shown to be related to the bending stiffness and mass of plate 3 on the air side.   These equivalencies provide a physical explanation for the relations \eqref{756} and \eqref{757}. 
{Summarizing the results obtained below:
\beq{6-6}
\begin{aligned}
\kappa_1 &= 720 \frac{D_1}{d^4}, \quad   \mu_1 = m_2 +\frac 12 m_1, 
\\
\kappa_2 &= 500 \frac{D_3}{d^4},  
\quad   \mu_2 = m_3.
\end{aligned}
\eeq
}

\subsection{Two resonances}\label{5.1}

The conditions for free vibration, or resonance, of the 3-plate model are derived and discussed in  \ref{A1}.  {It is shown there }that 
the exact condition, Eq.\ \eqref{300}, has a single zero  very close to the  zero for plate 3 alone, i.e. the plate 3 resonance condition  $S_{p3}=0$.  This simplifies to $g\big( (  m_3/ D_3\big)^{1/4} d  \sqrt{\omega},0\big)=0$ (see Eq.\ \eqref{452})
which is the same as the symmetric resonance frequency for a plate of length $d$ \cite[Ch.\ 11.5.2]{rao2019vibration}.   The first  positive zero of $g(\alpha ,0)$  is  $\alpha = 4.73$, implying  
\beq{202}
f_0  \approx \frac{3.56}{d^2} \sqrt{\frac{D_3}{m_3}}, 
\eeq
in agreement with \eqref{757} {and defining $\kappa_2$ and $\mu_2$ of \eqref{6-6}.}
The other resonator 
is plate 1 in combination with plate 2. Plate 1 acts mainly as a  spring ({it mostly bends, as shown in  the movies below}), and the central plate acts like a pure mass (in the movies, it mostly has a translational motion).  
Then, $\omega_0 \approx \sqrt{ \frac{ \kappa_1}{ m_\text{eff}} } $, 
where the equivalent stiffness follows from the quasi-static analysis of a flex-layer \cite{bakhtiary2024analytical,HBYANN2025},
$ \kappa_1 \approx  {720 D_1}/{d^4 } $ and $ m_\text{eff} \approx m_2$.  This expression for $\omega_0$ agrees with \eqref{756}.   It is shown in  \ref{appC} that  $ m_\text{eff} \approx m_2 + \frac 12 m_1$  provides a more accurate approximation for the effective mass. 
Assuming the plates are the same material, Eq.\ \eqref{202} and   
$\omega_0 \approx \sqrt{ \frac{ \kappa_1}{ m_\text{eff}} } $ (see Eq.\ 
\eqref{203}) provide a relation between $h_1$, $h_2$, and $h_3$: 
\beq{h1h2h3_v2}
h_3^2 \approx  \frac{1.44 \, h_1^3}{h_2 + 0.5 h_1} 
\eeq
For example: if $h_1 = 0.5$ mm and $h_2 = $ 8.376 mm, then $h_3$ is obtained 0.144 mm using Eq.\  \eqref{h1h2h3_v2}, while our data shows 0.14 mm.

The motion of the three-plate system is shown in the following videos for transmission at, approximately, $f_0 =500$ Hz, and $f_0 = 1,000$ Hz, {based on the examples above of case 1 in Table \ref {tab1} and Fig.\ \ref{4cases_E}, and also, of case 1 from  Table \ref{tab2} and Fig.\  \ref{4cases_E_1000Hz}.  In each case the operating frequency $f$ is taken slightly below or slightly above the central frequency $f_0$.}
\begin{enumerate}
\item 
\href{https://drive.google.com/file/d/1gex8wEHQ1KByfCOhlOJsSLGkDXMn4UWv/view?usp=sharing}{Flex-layer motion for  $f = 482$ Hz and $f_0 = 500$ Hz}.
\item 
\href{https://drive.google.com/file/d/1FxAGaGveQt1u3dItYCB2QGkhRZQnZm1R/view?usp=sharing}{Flex-layer motion for $f = 520$ Hz and $f_0 = 500$ Hz}.
\item 
\href{https://drive.google.com/file/d/1QxyGcnhG4a153Sb_QQa8nHfLezdMHgP6/view?usp=sharing}{Flex-layer motion for $f = 956$ Hz and $f_0 = 1,000$ Hz }.
\item 
\href{https://drive.google.com/file/d/1chHJ_MNF4yprFqt9tfchI-8sctR7uBog/view?usp=sharing}{Flex-layer motion for $f = 1,036$ Hz and $f_0 = 1,000$ Hz }.
\end{enumerate}
{It is clear from the videos that plates facing air and water}  oscillate out of phase at the frequency below $f_0$ and in phase above it.  This dynamic response is characteristic of a 2DOF system with closely spaced resonances.

\subsection{Two impedances}\label{5.2}

\begin{figure}[H]
    \centering
    \subfigure[]{\includegraphics[width=0.49\textwidth]{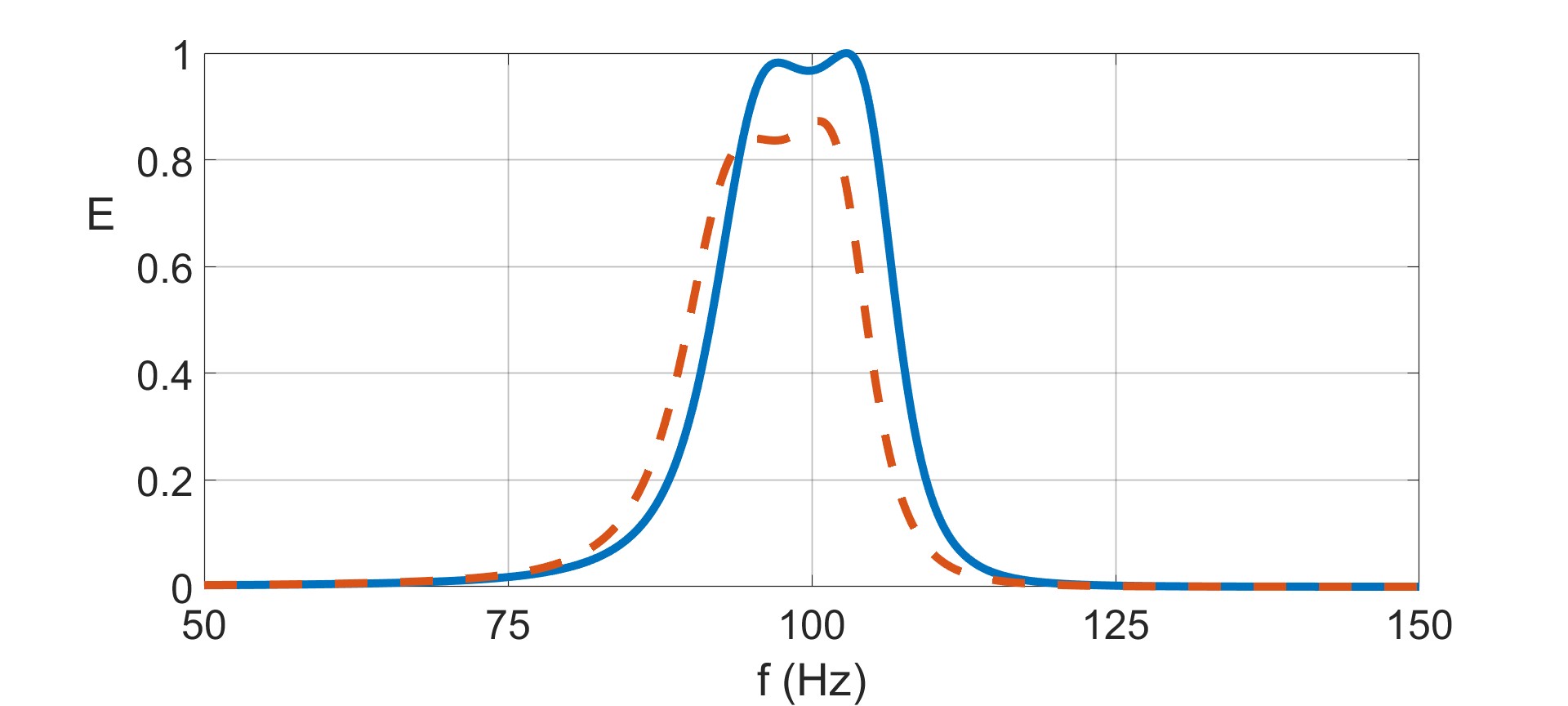}}
    \subfigure[]{\includegraphics[width=0.49\textwidth]{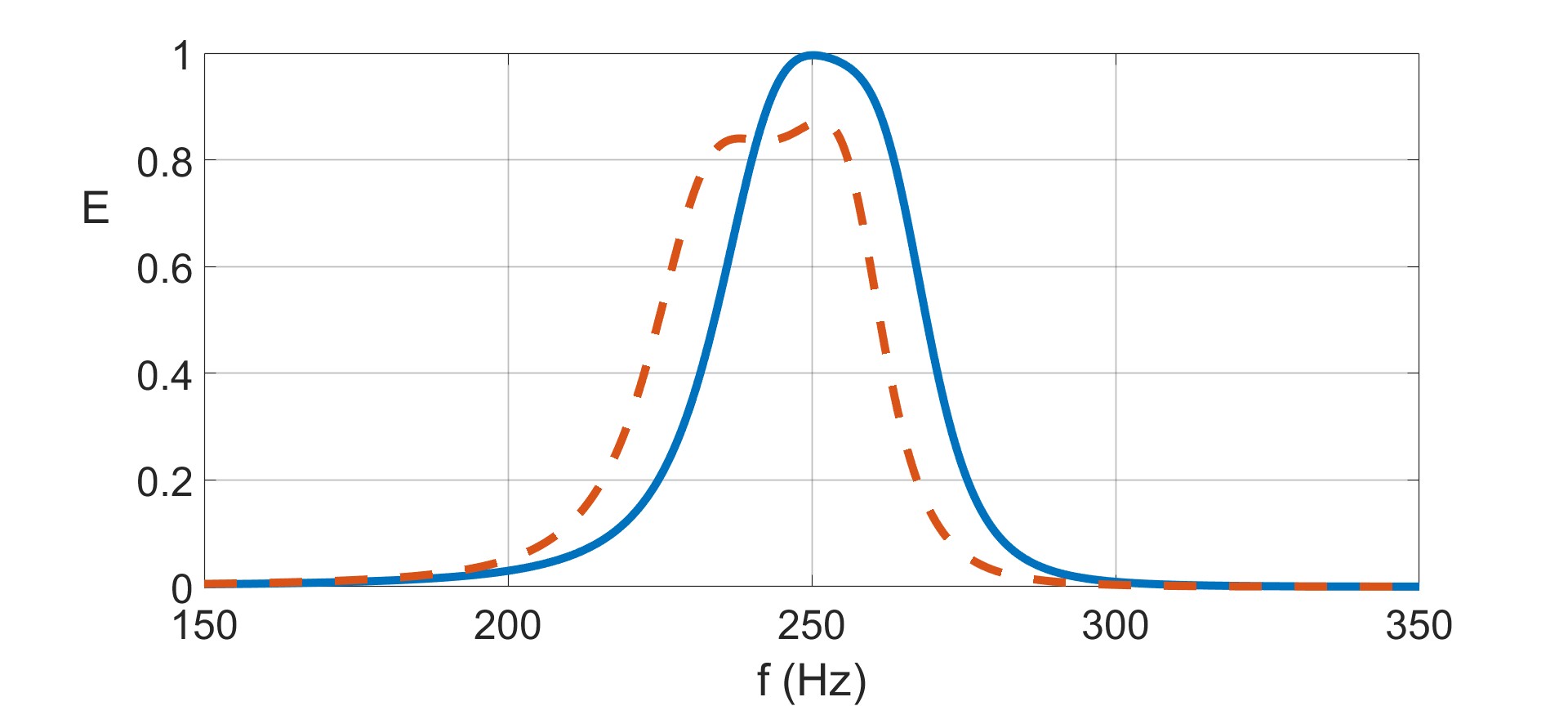}}
    \subfigure[]{\includegraphics[width=0.49\textwidth]{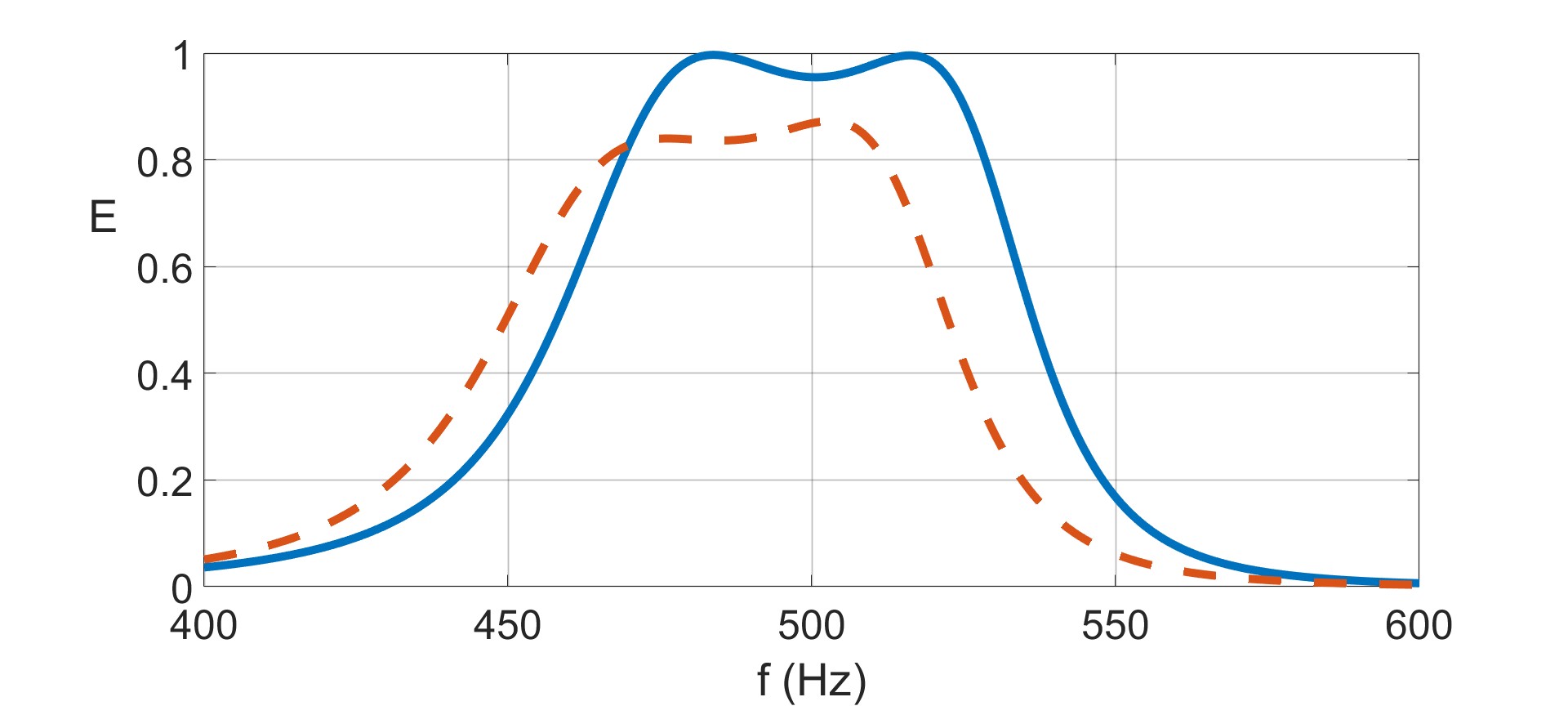}}
    \subfigure[]{\includegraphics[width=0.49\textwidth]{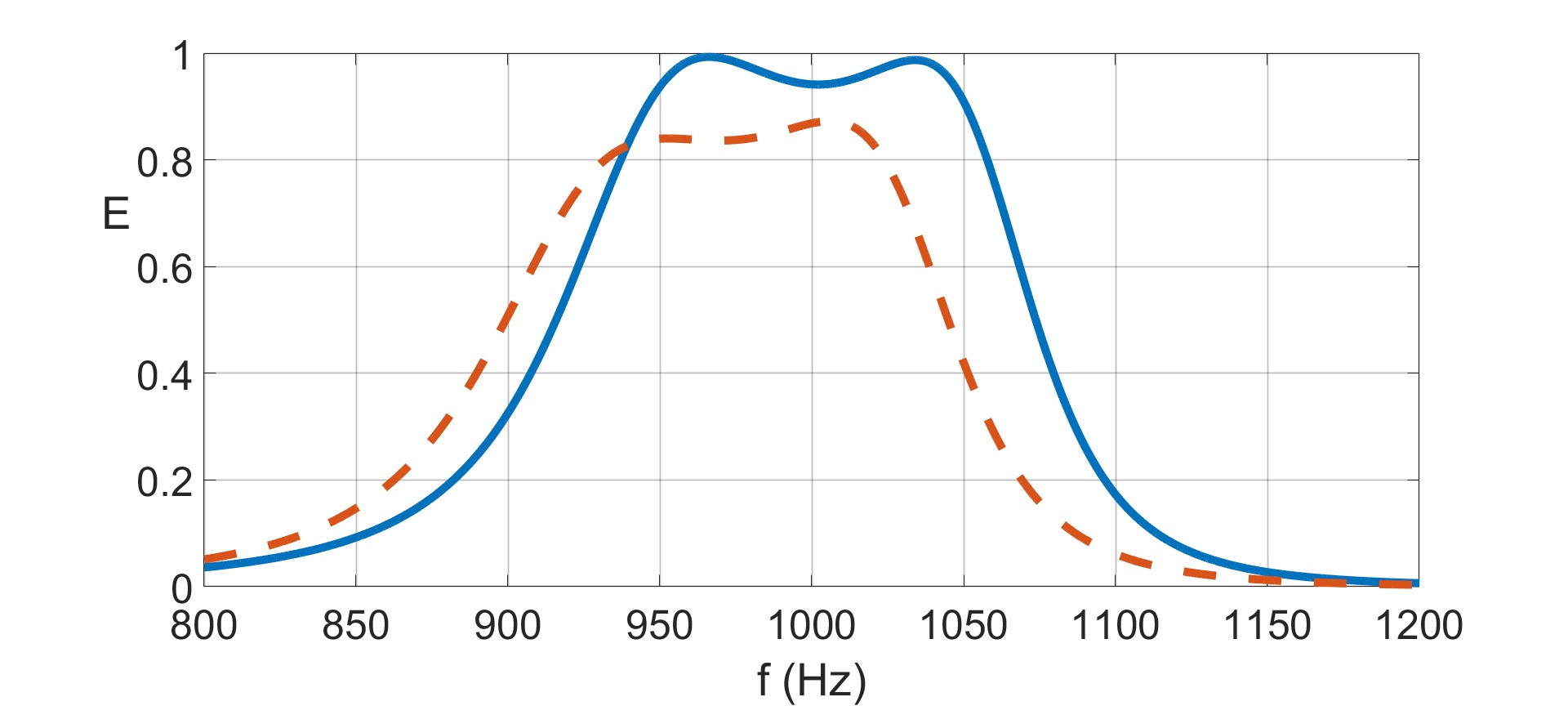}}
    \caption{Solid lines and dashed lines are respectively the energy transmission obtained using the optimization process and approximation equations (\eqref{446}, \eqref{354}, and \eqref{236}). The  approximate and final parameters are listed from Table \ref{tab5_h1h2h3d_comparison}. }
    \label{newEqcomp}
\end{figure}

We designate the resonances \eqref{756} and \eqref{757} as 1 and 2, i.e. resonance 1 is {at frequency} $\sqrt{ \frac{ \kappa_1}{ m_2 }} $ {(ignoring the mass correction $\frac 12 m_1$ of \eqref{6-6})} and  resonance 2 is at {$\sqrt{ \frac{ \kappa_2}{ m_3 } }$  with $\kappa_2$ defined in \eqref{6-6}}.  The associated impedances 
 $Z^{(1)} = \sqrt{   \kappa_1  m_2 }$ and 
{$Z^{(2)} = \sqrt{   \kappa_2  m_3 }$} may be written as 
\beq{945}
Z^{(1)} = \omega_0 m_2, \qquad 
Z^{(2)} = \omega_0 m_3. \qquad
\eeq
It then follows from Eq.\ \eqref{446}$_2$ that 
$Z^{(2)} / Z^{(1)} \approx \epsilon^{1/2}$, which is in agreement with the same ratio for the impedances defined in Eq.\ \eqref{033}.
Identifying the impedances \eqref{945} with those in \eqref{033} implies the relation
\beq{833}
\kappa_1 m_2 \approx Z_a^{1/2} Z_w^{3/2}.
\eeq
Assuming the plates are of the same density, and using Eq.\ \eqref{446} gives the alternative relations
\beq{944}
\kappa_1 m_3 \approx Z_aZ_w 
\quad \Leftrightarrow 
\quad
\kappa_1 m_1 \approx 0.886\, Z_a^{5/6} Z_w^{7/6}.
\eeq
This implies {a relation between $d$ and $h_2$ that does not involve frequency}
\beq{354}
d \approx   \bigg(
\frac{41.73 \,\rho_{s2} \  E}{ (1-\nu^2) Z_a^{-1/2} Z_w^{5/2}} 
\bigg)^{1/4}\, h_2.
\eeq
For aluminum we have  $d \approx 2.834\, h_2$.
For a given $f_0$,  $h_2$ follows from equating \eqref{945}$_1$ and \eqref{033}$_1$, 
\beq{236}
h_2 \approx \frac{ Z_a^{1/4} Z_w^{3/4} }{\rho_{s2} \  \omega_0 }.
\eeq
The other systems dimensions $h_1$, $h_3$ and $d$ can then be determined from Eqs.\ \eqref{446} and \eqref{354}.
{Table \ref{tab5_h1h2h3d_comparison} compares the  dimensions obtained from these approximations with numerically optimized values for four transmission frequencies.  The associated transmittivities  shown in Fig.\  \ref{newEqcomp} indicate the accuracy of the approximations. }

{One of the challenges that we faced when finding the optimal parameters was that, based on our simulations and optimization runs, generating a well-defined Pareto front typically requires several hours of computation. This difficulty comes from the large range of possible parameter values. By identifying approximately optimized parameters in equations (\eqref{446}, \eqref{354}, and \eqref{236}), we can use them as the initial population for the optimization process. As a result, the optimization converges much faster, and the final results are more refined.}

\begin{center} 
\begin{table}
\centering
\begin{tabular}{ |c|c|c|c|c| } 
\hline
Parameters & 100 Hz & 250 Hz & 500 Hz & 1000 Hz \\
\hline   \hline
\multirow{2}{4em}{$\tilde{h}_1$ (mm)\\${h_1}$ (mm)} 
&6.451 &2.580 &1.290 & 0.645\\
&6.786 &2.865 &1.342 & 0.714\\
\hline   \hline
\multirow{2}{4em}{$\tilde{h}_2$ (cm)\\${h_2}$ (cm)} 
&11.30 &4.521 &2.260 & 1.130\\
&10.48 &4.918 &1.979 & 1.027\\
\hline   \hline
\multirow{2}{4em}{$\tilde{h}_3$ (mm)\\${h_3}$ (mm)}
&1.847 &0.739 &0.369 & 0.184\\
&1.986 &0.786 &0.398 & 0.213\\
\hline   \hline
\multirow{2}{4em}{$\tilde d$ (cm)\\${d}$ (cm)} 
&32.03 &12.81 &6.40 & 3.20\\
&33.12 &13.15 &6.64 & 3.44\\
\hline
\end{tabular}
\caption {The approximated ($\bar h_1$ etc., obtained from Eqs.\ \eqref{446}, \eqref{354}, and \eqref{236}) and optimized values of $h_1$, $h_2$, $h_3$ and $d$. The associated transmittivities are  plotted in Fig.\ \ref{newEqcomp}.} \label{tab5_h1h2h3d_comparison}
\end{table}
\end{center}

\section{Summary and conclusions} \label{sec6} 

{We have presented a modification of the recently proposed flex-layer  transformer  \cite{HBYANN2025} that displays significantly improved transmission properties.  The present model, like the original  flex-layer design,  uses  purely solid materials to achieve impedance matching between water and air at a selected central frequency. We have shown that the bandwidth of the flex-layer  impedance transformer  \cite{HBYANN2025} can be significantly broadened by placing a mass-like plate between the air and water side plates, see Fig.\ \ref{Flex_3Plates_Optimal}. In particular, the lowest achievable Q-factor of the transmission resonance, which was found to be $Q_0 =\frac 1{2\sqrt{\epsilon}} = 30.59$ for the original model, 
becomes $Q \approx \sqrt{Q_0} = 5.53$, where  $\epsilon = Z_a/Z_w$ is the ratio of the air and water impedances $(\epsilon = 2.672 \times 10^{-4})$.
These results follow from a detailed analysis of the new flex-layer model as a two degree of freedom resonator in Section \ref{sec5}, and by comparison of the 2-DOF system with a binomial impedance transformer optimized for bandwidth, \ref{appB}. }

{The initial model considered in Fig.\ \ref{Flex_3Plates} contains several free parameters, such as the plate thicknesses and the rib spacing - four independent quantities, in addition to the choice of material properties (density and stiffness).  A major  goal of this paper has been to try to understand how the acoustic transmission performance depends upon this parameter space, and in the process to find specific parameter sets that provide simultaneously optimized bandwidth and transmittivity.  This ambitious objective has been met by first using the derived  analytical solution of Section \ref{sec2} to perform extensive numerical optimization experiments in Section \ref{sec3}.   The simulations indicate that   optimum transmission  is obtained if the center plate is far thicker than the ones facing water and air.   This means the center plate acts  as an effective mass,  which allows us to recast the three-plate transformer design  as a  two-plate model with an effective mass-like impedance between the plates, Fig.\ \ref{Flex_3Plates_Optimal}.  
This single observation, obtained by numerical means, allows us to use the 
much simpler two plate flex-layer design of \cite{HBYANN2025} with an added rib impedance to model the central mass. }

{The realization that the 3-plate model of Fig.\ \ref{Flex_3Plates} 
can be reduced to the 2-plate flex-layer that includes a mass-like rib impedance, a design that had actually been previously modeled \cite{HBYANN2025}, is perhaps the major takeaway from this article.  
This simplification enables us to characterize the transformer using asymptotic analysis based on the small parameter $\epsilon = Z_a/Z_w$.  The principal results are asymptotic approximations for the system dimensions.  Thus, 
the thickness $h_2$ of the central mass follows from Eq.\ \eqref{236}, and Eqs.\ 
\eqref{446} and \eqref{354} then yield $h_1$, $h_3$ and $d$ in terms of $h_2$: 
\beq{416}
h_2 \approx \frac{  Z_w \, \epsilon^{1/4} }{\rho_{s2} \ \omega_0 }, 
\quad
\frac{h_1}{h_2} \approx 0.886\,\epsilon^{1/3},
\quad
\frac{h_3}{h_2} \approx   \epsilon^{1/2},
\quad
\frac{d}{h_2} \approx  2.542 \, \Big( \frac{Z_p}{Z_w} \Big)^{1/2}\, 
\epsilon^{1/8}
\eeq
where 
$Z_p = \rho_s c_p$ is the plate impedance with $c_p = \sqrt{E/\rho_s(1-\nu^2)}$ the plate longitudinal wave speed.  For aluminum we have 
$d \approx 2.834\, h_2$.
These asymptotic approximations are not only interesting in their own right but they also serve as initial starting points for fast optimization using the analytical solution developed in Section \ref{sec2}.    We have found this to be extremely useful in speeding up numerical parameter searches. 
}

{The impedance transformer model considered here is, like the one studied previously  \cite{HBYANN2025}, a two dimensional design that assumes ribs that are infinitely long in the third dimension.  The system is also considered to be unbounded in the $y-$direction, allowing mathematical simplifications appropriate to periodic infinite systems.  Future work will examine  designs that are three-dimensional and are of finite extent. }

\section*{Appendix}
\appendix        

\section{Standing  wave resonances with and without fluid-loading}\label{appA}

\subsection{Exact dispersion relations}\label{A1}

Consider the two-plate flex-layer (plate 1 and plate 3) with no incident wave, in which case it follows from \cite[Eq.\ (4.7)]{HBYANN2025}
that 
\beq{99}
\begin{aligned}
    \big( \frac d{Z_{0+}} + S_1(\xi)  + S_3(\xi) \big) \, q_+ + 
   \big( S_3(\xi)  - S_1(\xi) \big) \, q_- &=0, 
    \\
     \big( S_3(\xi)  - S_1(\xi) \big) \, q_+ +    
 \big( \frac d{Z_{0-}} + S_1(\xi)  + S_3(\xi) \big) \, q_- &=0.
\end{aligned}
\eeq
Setting $ \frac 1{Z_{0-}} \to 0$ this implies that 
 $b(\xi) =0$, where $b$ is defined in Eq.\ \eqref{89}, is the condition for the existence of free waves of wavenumber $\xi$ along the flex-layer system under fluid loading \cite{Skelton2018}, that is 
 \beq{09}
 \frac 1{S_1(\xi)} + \frac 1{S_3(\xi)}  +\frac 1{s_{p2}}  = 0.
 \eeq
 
The related condition for the flex-layer without fluid loading is 
 \beq{091}
 \frac 1{S_{p1}(\xi)} + \frac 1{S_{p3}(\xi)}  +\frac 1{s_{p2}}  = 0
 \eeq
 where $S_{p1}$ and $S_{p3}$ are defined for the dry plates by 
 $ S_{pj}(\xi)  =  
 \sum_{m=-\infty}^\infty     \hat Y_{pj} \big(\xi +  \frac{2\pi m}{d}\big)$, 
 $j=1, 3$, 
 see Eq.\ \eqref{X}. 
Equation \eqref{091} can be simplified by using the identity
$S_{pj}(\xi) = -\frac{\ii \omega}{m_j} \beta_j^4 \, 
g\big(  \beta_j \sqrt{\omega}, \xi d \big) $, $j=1,3$, 
where $\beta_j = \big(  m_j/ D_j\big)^{1/4} d$
and 
\cite{Colquitt2017}
\beq{45}
g(\alpha,\zeta) \equiv 
\sum_n \frac 1 { (\zeta - 2\pi n)^4 - \alpha^4 } 
= \frac 1{4\alpha^3 } 
\bigg(
\frac{\sin \alpha} {\cos\alpha  - \cos \zeta} - 
\frac{\sinh \alpha} {\cosh\alpha  - \cos \zeta}
\bigg).
\eeq
The zeros of \eqref{091} correspond to traveling waves in the multi-plate system.   Zeros for $\xi=0$ represent cut-on frequencies, that are also resonances of the unit cell. 
Noting that 
\beq{452}
g(\alpha,0) = 
  - \frac 1{4\alpha^3 } 
\bigg(
\cot \frac{\alpha}2 +
\coth \frac{\alpha}2
\bigg).
\eeq
the condition for resonances is
\beq{300}
\sum_{j=1,3}
\frac{4m_j} {\beta_j \sqrt{\omega} }
\bigg(
\cot \frac{ \beta_j \sqrt{\omega} }2
 +
\coth\frac{ \beta_j \sqrt{\omega} }2
\bigg)^{-1}
+ m_2  = 0.
\eeq
A plot of the left member in \eqref{300} for the optimum solutions presented shows that it has a zero very close to the total transmission frequency.  For instance, for the example of Fig.\ \ref{E_verif}(b) with $h_1 = $ 1.508 mm,  $h_2 = $ 16.1 mm, $h_3 = $ 0.518 mm, $d = $ 7.60 cm, the zero is at $508$ Hz.  Similarly, the single zero is at $1007$ Hz for 
Case 1 of Fig.\  \ref{4cases_E_1000Hz}.  
In general, we find that the zero of  \eqref{300} is very close to that of  $S_{p3}$.  This means that the main factor in determining the full transmission frequency is the plate 3 resonance condition  $S_{p3}=0$.

The presence of a single zero is perhaps surprising in view of the well known prescription for a two-layer impedance transformer comprising two quarter-wavelength layers each with the resonance frequency of the transmission.  This two degree of freedom system is explored below in detail.  In particular, it is shown that the combined system has two resonant frequencies but only one is close to the transmission frequency, in agreement with our numerical observations for Eq.\ \eqref{300}.

\subsection{Approximate  dispersion relation}\label{A2}
The originator of acoustic impedance transformer theory, Hansell  \cite{Hansell}, recommended that each layer have (quarter-wavelength) resonance equal to the desired transmission frequency, in addition to specific values for the layer impedances.  Variations on this theme were developed in the middle of the 20$^{th}$ century in microwave applications \cite{Collin1955}.  The present model is analogous to a 2-layer transformer in that it has two degrees of freedom when viewed as a lumped parameter system.  This simple model replaces plates 1 and 3 with equivalent springs, 
\beq{3=}
 {\boxed{m_1}} \,\,\, \hlin  \overset{\kappa_1}{\spring}  {\boxed{m_2}}  \,\,\,  \hlin  \overset{\kappa_3}{\spring}  {\boxed{m_{3}}}   
\eeq
where $ \kappa_1 \approx  {720 D_1}/{d^4 } $  with a similar expression  for $\kappa_3$ {are quasistatic flexural approximations}  \cite{HBYANN2025}.
For fixed center of mass, this is a 2-degree of freedom system with modal frequencies satisfying
\beq{3-41}
\big(\omega^2- \omega^2_{01} \big) \big(\omega^2- \omega^2_{03} \big)
- \frac{m_1}{m_2}  \omega^2_{01} \big(\omega^2- \omega^2_{03} \big) 
-  \frac{m_3}{m_2}  \omega^2_{03} \big(\omega^2- \omega^2_{01} \big)= 0
\eeq
where $ \omega^2_{0j} = \kappa_j/m_j$, $j=1,3$.
In the optimal structures we find that $\omega_{03} \approx 
\omega_0$ while $\omega_{01} $ is several times larger.  More significant is the fact that $m_2$ is far larger than $m_1$ and $m_3$, implying modal frequencies
\beq{933}
 \omega^2_{j} \approx  \omega^2_{0j} 
  \big( 1+  \frac{m_j}{m_2} \big) , \quad j=1,3.
\eeq
Hence, $\omega_3 \approx \omega_0$ is the only zero near the transmission frequency, in agreement with the more complete model represented by the dispersion relation  Eq.\ \eqref{300}.

\section{Optimal two-layer impedance transformer}\label{appB}

\begin{figure}[H]
    \centering
    \includegraphics[width=0.6\textwidth]{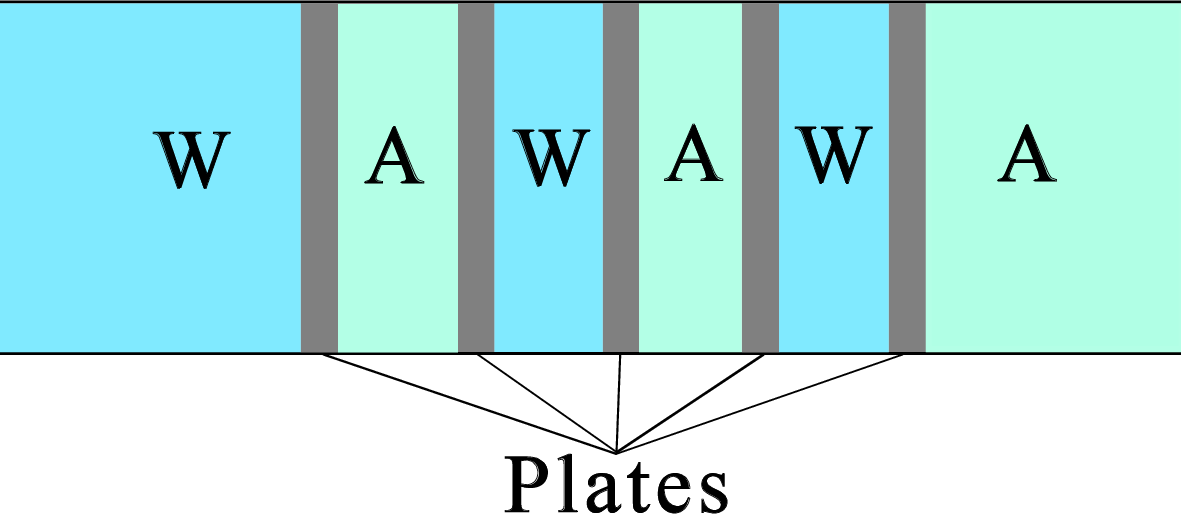}
    \caption{{A simple model of a sub-wavelength two layer graded impedance trasnsformer.}}
    \label{W-A-W-A-W-A}
\end{figure}

For the purpose of comparison with the flex-layer model we choose the   classical model of \cite{Hansell}, for which the  optimal system has layers with quarter-wavelength resonances equal to the transmission frequency and impedances 
\beq{033}
Z^{(1)} = Z_a^{1/4}Z_w^{3/4}, \quad 
Z^{(2)} = Z_a^{3/4}Z_w^{1/4}
\eeq
adjacent to water and air, respectively.  This is also known as the binomial transformer design \cite{JohnsonHansen}, \cite[pp.\ 272-273]{Southworth1950}, \cite[pp.\ 350-352]{Collin2001}, see \cite{Ahmed2021} for a review.
The Chebyshev  (or Tchebycheff) model \cite{Collin1955,Cohn1955}, \cite[pp.\ 352-360]{Collin2001} is an alternative design for  optimum bandwidth. It is however more complicated than the binomial model and 
does not yield  significantly different performance.    

We interpret the layers as spring-mass systems with stiffness 
$\kappa_j$ and mass $\mu_j$, $j=1,2$.  For a given transmission frequency $\omega_0$, it follows from 
\cite{Hansell} that 
$\omega_0^2 = \kappa_j/\mu_j$, $j=1,2$, and 
\eqref{033}  translates to 
\beq{232}
\begin{aligned}
 \kappa_1 = \epsilon^{-1/4}\, Z_e \,\omega_0, \qquad   \mu_1 &= \epsilon^{-1/4}\, \frac{Z_e}{ \omega_0},
\\
\kappa_2 = \epsilon^{1/4}\, Z_e \,\omega_0, \, \, \, \qquad   \mu_2 &= \epsilon^{1/4}\, \frac{Z_e}{ \omega_0},
\end{aligned}
\eeq
where $\epsilon = {Z_a/Z_w}$. 
The impedance facing the water is found by considering the system 
$\overset{ F \to}{\underset{v \to}{\Big|}} \hlin  \overset{\kappa_1}{\spring}  {\boxed{\mu_1}}  \,\,\,
\hlin  \overset{\kappa_2}{\spring}  {\boxed{\mu_2}}  \,\,\, \hlin \,\boxed{Z_a} $
 and yields 
\beq{88} 
\frac Fv \equiv Z_\text{eff} = \Big\{ -\frac { \ii\omega}{\kappa_1} +\Big[- \ii \omega \mu_1 +\Big( -\frac { \ii\omega}{\kappa_2} + \frac 1{Z_a - \ii \omega \mu_2} \Big)^{-1} \Big]^{-1}  \Big\}^{-1}.
\eeq
The {effective impedance then follows from  \eqref{232} and \eqref{88} as 
\beq{892}
{Z_\text{eff} }  
= \ii  \epsilon^{1/4} {Z_w}\, \bigg\{
\Omega  - \Big[ \Omega  
+ \epsilon^{1/2}  \Big(    \frac 1{  \Omega  +\ii  \epsilon^{1/4}} - \Omega 
\Big)^{-1} 
\Big]^{-1}
\bigg\}^{-1}  ,
\eeq}
where $\Omega = \frac{\omega}{\omega_0}$.  The reflection coefficient, $R = \frac{Z_\text{eff}-Z_w}{Z_\text{eff}+Z_w} $, is 
\beq{2377}
R = \frac
{  \epsilon  - \big(\Omega^2 -1\big)^2 }
{  \big(\Omega^2 -1 -\epsilon^{1/2} \big)^2 - 2 \epsilon^{1/2}
+ 2\ii  \epsilon^{1/4} \Omega  \big(\Omega^2 -1 -\epsilon^{1/2} \big) }
\eeq
and the transmitted energy is 
\beq{5==4}
\text{E} = 1 - |R|^2 = 
\frac 1{1 + \frac{\epsilon}4 \Big[ \frac{(\Omega^2 -1)^2}{\epsilon} -1 \Big]^2 }
.
\eeq
The above derivation has not used the fact that $\epsilon \ll 1$, which provides the simple and accurate asymptotic approximation
\beq{5==41}
E \approx \Big(1 + \frac 4{\epsilon} \big( \Omega-1  \big)^4   \Big)^{-1}.
\eeq

The $Q$-factor follows from   \eqref{5==41} as $Q\approx \frac 1{\sqrt{2}\,\epsilon^{1/4}}$ which is the square root of the $Q$-factor for the optimized single layer transformer \cite{HBYANN2025}. 
For air and water 
$\epsilon = 2.672 \times 10^{-4} $ implying the optimal $Q \approx 5.53$ {for the flex-layer model considered here.   We consider this $Q-$factor as {\em optimal} because it corresponds to the 
binomial transformer 2-layer design mentioned above. The physical origin of the optimal $Q$ can be attributed to radiation damping alone.}

The stiffnesses and masses of \eqref{232} can be realized, in principle if not in practice, by thin layers of air and water \cite{HBYANN2025}.  
The transformer configuration with water on the left and air on the right is 
 \textcolor{blue}{w}\,\textcolor{red}{{\vrule width 1pt}}\,a$_1$\,\textcolor{red}{{\vrule width 1pt}}\,\textcolor{blue}{w}$_1$
 \textcolor{red}{{\vrule width 1pt}}\,a$_2$\,\textcolor{red}{{\vrule width 1pt}}\,\textcolor{blue}{w}$_2$\textcolor{red}{{\vrule width 1pt}}\,a
 where the    \textcolor{red}{{\vrule width 1pt}}  indicate thin plates or membranes separating the air and water. 
The {air layers}, a$_1$ and a$_2$, act as compressible springs while the water layers, \textcolor{blue}{w}$_1$ and \textcolor{blue}{w}$_2$ and the separators all act as masses.  Let the 
air and water thicknesses be
$d_{aj}$ and $d_{wj}$, $j=1,2.$
Since the effect of the  separators is to reduce $d_{wj}$ and leave 
$d_{aj}$ unchanged, we ignore them for simplicity.  The air and water thicknesses then satisfy \cite{HBYANN2025} 
\beq{2}
d_{aj} = \frac{\rho_a c_a^2}{\kappa_j}, 
\qquad 
d_{wj} = \frac{\mu_j}{\rho_w} , \quad j=1,2.
\eeq
  \begin{figure}[H] 
  \centering
    \includegraphics[width=0.6\textwidth]{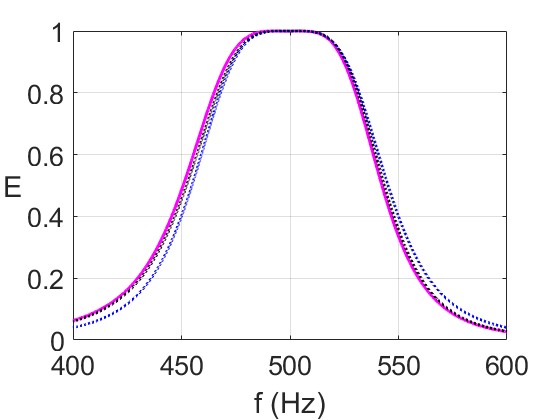}
  \caption{ Transmitted energy for unit incident energy  from the water side, $f_0=500$ Hz. The solid curve is a full wave   simulation and the dashed black curve is the lumped-parameter model \eqref{5==4}.  The   dashed blue curve 
  is the $\epsilon$-asymptotic approximation \eqref{5==41}.  
  }
  \label{fig99}
   \end{figure}  
   
For instance, at $f_0 = 500$ Hz we find
$d_{a1}= 0.222$ mm,  $d_{w1}= 61.05$ mm,   $d_{a2}=13.59$ mm,  $d_{w2}=    0.998$ mm.
The very thin nature of $d_{a1}$ and $d_{w2}$ makes this  hard to imagine as practical \cite{HBYANN2025}. However, the 
  \textcolor{red}{{\vrule width 1pt}}\,a$_1$\,\textcolor{red}{{\vrule width 1pt}}\,\textcolor{blue}{w}$_1$
 \textcolor{red}{{\vrule width 1pt}}\,a$_2$\,\textcolor{red}{{\vrule width 1pt}}\,\textcolor{blue}{w}$_2$\textcolor{red}{{\vrule width 1pt}}  transformer 
 serves as an instructive comparison to the flex-layer model.  Thus, 
 Fig.\  \ref{fig99} compares the full wave simulation for this transformer  with the lumped parameter model prediction of Eq.\ \eqref{5==4}, with almost perfect agreement.   Figure \ref{fig99} also shows the asymptotic approximation   \eqref{5==41}. 
   

\section{Effective mass}\label{appC}

The approximation $\omega_0 \approx \sqrt{ \frac{ \kappa_1}{ m_2 }} $ 
of Section \ref{sec5} 
can be improved by taking into account the inertia of plate 1. 
By using the mode shape function $w_1(y) \approx \cos \big( \frac{2\pi}{d} y \big)$  for the first plate \cite{HBYANN2025}, we can find the equivalent mass of the first plate by assuming $w_1(y,t) = w_1(y) z(t)$:
\beq{203_meq}
\text{Kinetic Energy} = \frac{1}{2} \rho_s h_1 \dot{z}^2 \int_{-d/2}^{d/2} w_1^2(y)  
\,dy\ = \frac{1}{2} \left( 0.5 \rho_s h_1\right) \dot{z}^2 , 
\eeq
and therefore
$ m_\text{eff} \approx m_2 + 0.5 m_1 $.
\eeq
  Hence, $\omega_0 \approx \sqrt{ \frac{ \kappa_1}{ m_\text{eff}} } $ implies 
\beq{203}
 f_0 \approx 
 \frac{4.27}{d^2} \sqrt{\frac{D_1}{m_2 +\frac 12 m_1}}.
\eeq
Using the data for   case 1 in Table \ref{tab3_h1h2h3d} , Eqs. \eqref{202} and \eqref{203} result in 504.9 Hz and 524.8 Hz, respectively. On the other hand, for  case 1  in Table  \ref{tab4_h1h2h3d}, Eqs. \eqref{202} and \eqref{203} give 935.19 Hz and 1033.8 Hz, respectively. These results provide a good explanation for the two peaks that we observe for each case in Figs \ref{f0_500Hz_h1_h2_h3_d} and \ref{f0_1000Hz_h1_h2_h3_d}.


\end{document}